\documentclass[aps, reprint, superscriptaddress, floatfix, showpacs, amsmath, amssymb]{revtex4-1}

\usepackage{graphicx,color}
\usepackage{dcolumn}
\usepackage{bm}
\usepackage{hyperref}
\usepackage{lineno}
\usepackage{rotating}

\begin{document}

\title{Precision measurement of the longitudinal double-spin asymmetry for dijet production at intermediate pseudorapidity in polarized $pp$ collisions at $\sqrt{s}$ = 200 GeV}
\affiliation{Academia Sinica}
\affiliation{Abilene Christian University, Abilene, Texas   79699}
\affiliation{AGH University of Krakow, FPACS, Cracow 30-059, Poland}
\affiliation{Argonne National Laboratory, Argonne, Illinois 60439}
\affiliation{American University in Cairo, New Cairo 11835, Egypt}
\affiliation{Ball State University, Muncie, Indiana, 47306}
\affiliation{Brookhaven National Laboratory, Upton, New York 11973}
\affiliation{University of Calabria \& INFN-Cosenza, Rende 87036, Italy}
\affiliation{University of California, Berkeley, California 94720}
\affiliation{University of California, Davis, California 95616}
\affiliation{University of California, Los Angeles, California 90095}
\affiliation{University of California, Riverside, California 92521}
\affiliation{Central China Normal University, Wuhan, Hubei 430079 }
\affiliation{University of Illinois at Chicago, Chicago, Illinois 60607}
\affiliation{Chongqing University, Chongqing, 401331}
\affiliation{Creighton University, Omaha, Nebraska 68178}
\affiliation{Czech Technical University in Prague, FNSPE, Prague 115 19, Czech Republic}
\affiliation{Technische Universit\"at Darmstadt, Darmstadt 64289, Germany}
\affiliation{National Institute of Technology Durgapur, Durgapur - 713209, India}
\affiliation{ELTE E\"otv\"os Lor\'and University, Budapest, Hungary H-1117}
\affiliation{Frankfurt Institute for Advanced Studies FIAS, Frankfurt 60438, Germany}
\affiliation{Fudan University, Shanghai, 200433 }
\affiliation{Guangxi Normal University, Guilin, 541004}
\affiliation{University of Heidelberg, Heidelberg 69120, Germany }
\affiliation{University of Houston, Houston, Texas 77204}
\affiliation{Huzhou University, Huzhou, Zhejiang  313000}
\affiliation{Indian Institute of Science Education and Research (IISER), Berhampur 760010 , India}
\affiliation{Indian Institute of Science Education and Research (IISER) Tirupati, Tirupati 517507, India}
\affiliation{Indian Institute Technology, Patna, Bihar 801106, India}
\affiliation{Indiana University, Bloomington, Indiana 47408}
\affiliation{Institute of Modern Physics, Chinese Academy of Sciences, Lanzhou, Gansu 730000 }
\affiliation{University of Jammu, Jammu 180001, India}
\affiliation{Kent State University, Kent, Ohio 44242}
\affiliation{University of Kentucky, Lexington, Kentucky 40506-0055}
\affiliation{Lanzhou University}
\affiliation{Lawrence Berkeley National Laboratory, Berkeley, California 94720}
\affiliation{Lehigh University, Bethlehem, Pennsylvania 18015}
\affiliation{Max-Planck-Institut f\"ur Physik, Munich 80805, Germany}
\affiliation{Michigan State University, East Lansing, Michigan 48824}
\affiliation{National Institute of Science Education and Research, HBNI, Jatni 752050, India}
\affiliation{National Cheng Kung University, Tainan 70101 }
\affiliation{Nuclear Physics Institute of the CAS, Rez 250 68, Czech Republic}
\affiliation{The Ohio State University, Columbus, Ohio 43210}
\affiliation{Panjab University, Chandigarh 160014, India}
\affiliation{Purdue University, West Lafayette, Indiana 47907}
\affiliation{Rice University, Houston, Texas 77251}
\affiliation{Rutgers University, Piscataway, New Jersey 08854}
\affiliation{University of Science and Technology of China, Hefei, Anhui 230026}
\affiliation{South China Normal University, Guangzhou, Guangdong 510631}
\affiliation{Sejong University, Seoul, 05006, South Korea}
\affiliation{Shandong University, Qingdao, Shandong 266237}
\affiliation{Shanghai Institute of Applied Physics, Chinese Academy of Sciences, Shanghai 201800}
\affiliation{Southern Connecticut State University, New Haven, Connecticut 06515}
\affiliation{State University of New York, Stony Brook, New York 11794}
\affiliation{Instituto de Alta Investigaci\'on, Universidad de Tarapac\'a, Arica 1000000, Chile}
\affiliation{Temple University, Philadelphia, Pennsylvania 19122}
\affiliation{Texas A\&M University, College Station, Texas 77843}
\affiliation{Texas Southern University}
\affiliation{University of Texas, Austin, Texas 78712}
\affiliation{Tsinghua University, Beijing 100084}
\affiliation{University of Tsukuba, Tsukuba, Ibaraki 305-8571, Japan}
\affiliation{University of Chinese Academy of Sciences, Beijing, 101408}
\affiliation{United States Naval Academy, Annapolis, Maryland 21402}
\affiliation{Valparaiso University, Valparaiso, Indiana 46383}
\affiliation{Variable Energy Cyclotron Centre, Kolkata 700064, India}
\affiliation{Warsaw University of Technology, Warsaw 00-661, Poland}
\affiliation{Wayne State University, Detroit, Michigan 48201}
\affiliation{Wuhan University of Science and Technology, Wuhan, Hubei 430065}
\affiliation{Yale University, New Haven, Connecticut 06520}

\author{B.~E.~Aboona}\affiliation{Texas A\&M University, College Station, Texas 77843}
\author{J.~Adam}\affiliation{Czech Technical University in Prague, FNSPE, Prague 115 19, Czech Republic}
\author{L.~Adamczyk}\affiliation{AGH University of Krakow, FPACS, Cracow 30-059, Poland}
\author{I.~Aggarwal}\affiliation{Panjab University, Chandigarh 160014, India}
\author{M.~M.~Aggarwal}\affiliation{Panjab University, Chandigarh 160014, India}
\author{Z.~Ahammed}\affiliation{Variable Energy Cyclotron Centre, Kolkata 700064, India}
\author{E.~C.~Aschenauer}\affiliation{Brookhaven National Laboratory, Upton, New York 11973}
\author{S.~Aslam}\affiliation{Indian Institute Technology, Patna, Bihar 801106, India}
\author{J.~Atchison}\affiliation{Abilene Christian University, Abilene, Texas   79699}
\author{V.~Bairathi}\affiliation{Instituto de Alta Investigaci\'on, Universidad de Tarapac\'a, Arica 1000000, Chile}
% ...

\collaboration{STAR Collaboration}\noaffiliation

\date{\today}
\begin{abstract}
The STAR Collaboration reports precise measurements of the longitudinal double-spin asymmetry, $A_{LL}$, for dijet production with at least one jet at intermediate pseudorapidity $0.8 < \eta_{\rm jet} < 1.8$ in polarized proton-proton collisions at a center-of-mass energy of 200 GeV. This study explores partons scattered with a longitudinal momentum fraction ($x$) from 0.01 to 0.5, which are predominantly characterized by interactions between high-$x$ valence quarks and low-$x$ gluons. The results are in good agreement with previous measurements at 200 GeV with improved precision and are found to be consistent with the predictions of global analyses that find the gluon polarization to be positive. In contrast, the negative gluon polarization solution from the JAM Collaboration is found to be strongly disfavored.
\end{abstract}

\pacs{}

\maketitle


\section{Introduction\label{sec:Introduction}}

Understanding how the intrinsic spins of quarks and gluons, along with their orbital angular momenta, combine to yield the proton spin of $\hbar/2$ has been a major challenge in Quantum Chromodynamics (QCD)~\cite{Jaffe:1989jz}. Over the last twenty years, significant progress has been made towards unraveling the complexities of the proton spin structure, fueled by increasingly precise measurements from polarized deep inelastic lepton-nucleon scattering and hadronic proton-proton ($pp$) collision experiments. These experimental findings, integrated with sophisticated theoretical frameworks, have shed light on the spin structure of the proton. It has been established that the spins of quarks $(\Delta \Sigma)$ contribute approximately $30\%$ to the overall proton spin. Spins of gluons $(\Delta G)$ are found to make a substantial contribution of around $40\%$ in regions where the longitudinal momentum fraction ($x$) is larger than $0.05$. With these, the remaining contribution has been attributed to low momentum fraction gluons and the orbital angular momentum of partons within the nucleon (\cite{ref:NNPDF, ref:DSSV2014} and references therein).\par

Longitudinally polarized $pp$ collisions at the Relativistic Heavy Ion Collider (RHIC)~\cite{Alekseev:2003sk} have provided unique data to unravel the intricate spin structure of the proton with various final state observables in the experiment. The STAR and PHENIX collaborations have been instrumental in this quest, conducting key measurements of the longitudinal double-spin asymmetry, $A_{LL}$, for both inclusive jet~\cite{ref:STAR_2006_ALL,ref:STAR_2007_ALL,ref:STAR_2012pub_ALL,ref:Adamczyk_2015_ALL} and $\pi^{0}$ production in $pp$ collisions at $\sqrt{s}$ = 200 GeV~\cite{ref:STAR_2009_pi0_ALL,ref:STAR_2006_Endcap_pi0_ALL, PHENIX:2008sgl, PHENIX:2008swq, PHENIX:2014gbf}. These observations, when incorporated with data from inclusive and semi-inclusive lepton-proton scattering, placed substantial constraints on the DSSV14~\cite{ref:DSSV2014} and NNPDFpol1.1~\cite{ref:NNPDF} next-to-leading order (NLO) perturbative QCD global analyses.\par
 
In contrast to the inclusive observables, dijets measurement has potential to provide a more precise constraint on the kinematics of the parton interaction. Therefore, they could be a more effective tool to extract the momentum fraction dependence of the gluon helicity distribution~\cite{deFlorian:2009fw}. At leading order (LO), the dijet invariant mass ($M$) is directly proportional to the square root of the product of the partonic momentum fractions, encapsulated by the equation $M = \sqrt{sx_{1}x_{2}}$, and the sum of the pseudorapidities of the jet pair, $\eta_{3} + \eta_{4} = \mathrm{ln}(x_{1}/x_{2})$ (Here, it is $1 + 2 \rightarrow 3 + 4$). Investigating various dijet production topologies enables the examination of partonic interactions across both symmetric ($x_{1} = x_{2}$) and asymmetric ($x_{1} > x_{2}$ or vice versa) momentum fraction scenarios. Although it is not as simple as this in NLO, the dijet measurements allow us to better constrain the kinematics of the initial partons. Using the 2009 dataset, STAR reported the first measurement of the differential cross section and $A_{LL}$ for dijets near mid-rapidity in longitudinally-polarized proton-proton collisions at $\sqrt{s}=200$ GeV~\cite{ref:STAR_2009_Barrel_dijet_ALL}. The measured cross section agreed with NLO perturbative QCD expectations, and the extracted spin asymmetries exhibited good agreement with predictions from global analyses. STAR also presented the first measurement of $A_{LL}$ for dijet production with one or both jets entering the intermediate pseudorapidity region $0.8 < \eta_{\mathrm{jet}} < 1.8$, utilizing data collected in 2009 at $\sqrt{s} = 200 $ GeV~\cite{ref:STAR_2009_Endcap_ALL}. The reweighted fit results significantly improve our knowledge of gluon polarization in the region of $x > 0.1$. The representative value and 1-$\sigma$ uncertainty for truncated moments of the gluon helicity density changed from $0.133 \pm 0.035$ to $0.126 \pm 0.023$~\cite{DeFlorian:2019xxt}.\par

In 2012 and 2013, STAR recorded longitudinally polarized proton-proton collisions at $\sqrt{s} =$ 510 GeV with 82~pb$^{-1}$ and 250 pb$^{-1}$ respectively. Both inclusive jet and dijet $A_{LL}$ measurements were extended to higher collision energies, enhancing sensitivity to lower momentum fraction partons. These results~\cite{ref:STAR_2012_ALL, ref:STAR_2013_ALL}, presenting excellent agreement with measurements at $\sqrt{s} =$ 200 GeV in the overlapping $x$ region, offer crucial new insights into both the magnitude and shape of gluon polarization within the range $0.015 < x < 0.2$. Moreover, STAR has extended its investigations of $\sqrt{s}$ = 510 GeV $pp$ collisions to include $A_{LL}$ in $\pi^{0}$ production at forward rapidities~\cite{ref:STAR_FMS_pi0_ALL}, which provide sensitivity of $x$ down to $0.001$. PHENIX has also measured $A_{LL}$ for mid-rapidity $\pi^{0}$~\cite{PHENIX:2015fxo}, charged pions~\cite{ref:PHENIX_chargedPion_ALL}, and direct photons~\cite{ref:PHENIX_direct_photon_ALL} in $pp$ collisions at $\sqrt{s}$ = 510 GeV, enriching our understanding of the proton's internal spin dynamics.\par

STAR concluded the longitudinally polarized program in 2015 with $pp$ collisions at 200 GeV. The inclusive jet and dijet results at mid-rapidity~\cite{ref:STAR_2015results_ALL} confirm previous findings with twice more integrated luminosity.\par

In 2022, the JAM Collaboration conducted a global QCD analysis of the spin-dependent parton distribution functions, utilizing the inclusive jet data from RHIC. Remarkably, that analysis, which did not enforce the positivity constraint, revealed that two distinct solutions for the gluon polarization of opposite sign could provide nearly identical results to the experimental spin-dependent observables, $A_{LL}$~\cite{ref:jam_negative_gluon}. Nevertheless, more recent analysis from the DSSV group indicated that the scenario involving negative gluon polarization is disfavored by the data from both the direct photon and dijet productions~\cite{ref:dssv_spin2023}. The comparison of the direct photon result from PHENIX~\cite{ref:PHENIX_direct_photon_ALL} disfavored the negative solution at a confidence level of 2.8$\sigma$. Recently, the JAM Collaboration also found that the negative gluon helicity solution cannot simultaneously account for high-$x$ polarized DIS data along with lattice QCD calculations and polarized jet data~\cite{ref:Hunt-Smith:2024khs}.\par

In this paper, the STAR Collaboration presents a precise measurement of the longitudinal double-spin asymmetry for dijet production at intermediate pseudorapidity in $pp$ collisions at $\sqrt{s} =$ 200 GeV, utilizing data collected in 2015 with approximately twice the integrated luminosity compared to 2009. The results are measured as a function of dijet invariant mass and extend the sensitivity to different kinematic regions beyond midrapidity, thereby contributing valuable insights to the understanding of polarized hadronic interactions. A direct comparison is also made to the global analysis from JAM Collaboration, offering another test of the negative gluon polarization result.\par

The remaining parts of this paper are arranged as follows: Section~\ref{sec:Experiment} details the STAR detector subsystems relevant to this measurement. Section~\ref{sec:DataSim} describes the data sets and simulation samples. Section \ref{sec:Jetdijet} presents the jet reconstruction and dijet selection. Section \ref{sec:Methods} provides details on the experimental methods. The double-spin asymmetry $A_{LL}$ and the associated bias and uncertainties are discussed in Sec.~\ref{sec:asymmetry}. Section~\ref{sec:Results} presents the results. Finally, Sec.~\ref{sec:Conclusion} concludes.\par

\section{The STAR detector at RHIC}\label{sec:Experiment}

RHIC~\cite{Alekseev:2003sk} at Brookhaven National Laboratory stands as the world's first and only accelerator with the capability to collide the longitudinally or transversely polarized proton beams. Furthermore, RHIC is capable of colliding beams at both $\sqrt{s} =$ 200 and 510 GeV. In the 2015 polarized $pp$ running period, each beam was filled with 111 bunches of vertically polarized protons. Spin rotator magnets, located at each side of the STAR interaction region, manipulated the proton spin orientation from vertical to longitudinal and vice versa, enabling collisions of longitudinally polarized beams. To minimize potential spin-dependent systematic effects that could emerge from variations across different bunches, a distinct spin pattern was assigned to each proton bunch. The determination of beam polarization at RHIC is achieved through the use of two distinct types of polarimeters: the proton-carbon (pC) Coulomb-nuclear interference polarimeters and the polarized atomic hydrogen jet (H-jet) polarimeter~\cite{ref:pCPolarimeter,ref:RHIC_HJet}. The pC polarimeters are designed for fast measurements and are capable of measuring the relative polarization of the beam multiple times during a storage period, which typically spans eight hours. Meanwhile, the H-jet polarimeter provides an absolute measurement of beam polarization, serving as a calibration benchmark for the results obtained from the pC polarimeters. This dual-measurement approach ensures high accuracy and reliability in assessing the polarization states of the proton beams.\par

The Solenoidal Tracker at RHIC (STAR)~\cite{STAR:2002eio} is a large acceptance detector which has been operational for over two decades. The central part of the STAR detector is the Time Projection Chamber (TPC)~\cite{ref:tpcNIM}, which measures the momenta of the charged particles. The TPC operates within a magnetic field of 0.5 Tesla, effectively covering a broad range of pseudorapidity, $|\eta| < 1.3$. Complementing the TPC are the Barrel and Endcap Electromagnetic Calorimeters (BEMC and EEMC)~\cite{Beddo:2002zx, Allgower:2002zy}. BEMC has similar coverage as TPC while EEMC extends the kinematic reach of the BEMC to $1.09 < \eta < 2$ with full azimuth. These calorimeters serve dual roles, not only measuring the electromagnetic energy depositions, but also acting as triggering detectors for high-energy particles and jets. The vertex position detector (VPD)~\cite{ref:vpdNIM} and the zero-degree calorimeters (ZDC)~\cite{ref:RHIC_ZDC} are pairs of far-forward detectors, providing measurements of the relative luminosities of colliding bunches between different helicity states.\par

\section{Data and simulation}\label{sec:DataSim}
\subsection{Data sets and event selection}

The data analyzed in this study were collected by the STAR experiment in 2015 from longitudinally polarized $pp$ collisions at $\sqrt{s}$ = 200 GeV, with an integrated luminosity of 52 pb$^{-1}$. The luminosity-weighted polarizations for the two beams, known as the blue beam ($P_{\mathrm{B}}$, going to $+z$) and the yellow beam ($P_{\mathrm{Y}}$, going to $-z$),
were measured to be 53\% and 57\%, respectively. The relative uncertainty in the product of the beam polarizations, $P_{\mathrm{B}}P_{\mathrm{Y}}$, was determined to be 6.1\%~\cite{ref:RHICPolG}.\par

The event triggers employed for this analysis were jet patch (JP) triggers, which impose thresholds on the total transverse energy ($E_{\mathrm{T}}$) observed within fixed $\Delta \eta \times \Delta \phi = 1 \times 1$ regions of the BEMC and EEMC. The coverage area for these triggers spanned $-1 < \eta < 2$, organized into thirty distinct jet patches. This arrangement is comprised of five partially overlapping jet patches in $\eta$ in each six non-overlapping sectors in $\phi$. During the operational period of 2015, two levels of JP triggers were utilized to optimize the detection and recording of relevant collision events: JP1 and JP2. The JP1 trigger was set with a threshold of 5.4 GeV, designed to capture lower-energy events, while the JP2 trigger had a higher threshold of 7.3 GeV, targeting events with more substantial energy depositions. To efficiently manage the data-acquisition bandwidth and prioritize the recording of high-value data, all events that activated the JP2 trigger were recorded. Conversely, events that only activated the JP1 trigger were subject to prescaling. This strategy allowed for a balanced data collection, ensuring that events across a spectrum of energy depositions were sampled while maintaining a manageable volume of data for subsequent analyses.\par

\subsection{Embedded simulation samples}\label{sec:embedding}

Monte Carlo (MC) simulations are critical for correcting detector-related distortions in measured observables and for evaluating systematic uncertainties that might affect the results. In this analysis, simulated events were generated using \textsc{Pythia} 6.4.28~\cite{ref:Pythia6} with the Perugia 2012 tune~\cite{ref:PerugiaTunes}. A specific adjustment in the parameter PARP(90), which controls the energy dependence of the low-$p_{\mathrm{T}}$ cut-off for the underlying event (UE) generation process, was made from its default value of 0.24 to 0.213. This modification was made to align the simulated events with previous STAR measurements of $\pi^{\pm}$ yields in 200 GeV $pp$ collisions~\cite{ref:starPion2005,ref:starPion2012}. The application of this specific setup in \textsc{Pythia} has proven to be highly effective in describing various aspects of collision events, including jet production, hadronization within jets, and characteristics of the underlying event across both 200 and 510 GeV $pp$ collisions~\cite{ref:STAR_2012_ALL,ref:STAR_2013_ALL,ref:STAR_2015results_ALL,ref:STAR_invariant_mass,ref:STAR_groomed_jet,ref:STAR_UE_paper,ref:STAR_2012_2015_Collins}. To account for detector effects on the measurement, the generated events were processed through comprehensive detector simulations using \textsc{Geant}3~\cite{ref:Geant}, which incorporated the specific configurations of the 2015 STAR detector setup. To account for the complexities of actual experimental conditions, such as event pile-up and beam background effects, the simulated events were embedded into real `zero-bias' events from the 2015 data. These zero-bias events are randomly selected, independent of collision events, to capture background conditions and represent the operational environment of the data collection period.
By embedding the simulated events into these real-event backgrounds, the simulations more accurately replicate the conditions under which the experimental data were collected, ensuring that the analysis reflects the true performance of the detector and the real conditions of the experiment.\par

The simulation software also records the initial partonic hard scattering, the subsequent stages of fragmentation and hadronization where partons transform into hadrons, and the interaction of these final-state particles with the detector. These processes are categorized into three distinct levels of information: parton-level, particle-level, and detector-level, each offering unique insights into the collision events. Parton-level focuses on the initial conditions and the products of the partonic hard scattering, before any fragmentation or hadronization processes occur and excluding underlying event contributions. It provides a direct view into the theoretical predictions at leading order and allows comparisons with the calculations from perturbative QCD. After the initial scattering and parton showers, partons undergo fragmentation and hadronization, leading to the production of stable or long-lived particles at the particle-level. The detector-level replicates the detector's response to these particles, producing output in the same format as that generated by real experimental data. This multilevel approach facilitates a systematic analysis that ranges from theoretical calculations at the parton level, through the physical processes leading to stable particle formation, to the practical challenges of measuring these particles with a detector. It enables the precise correction of detector effects, the estimation of systematic uncertainties, and the refinement of theoretical models, thereby enhancing the accuracy and reliability of the experimental results.\par

\section{Jet reconstruction and dijet selection}\label{sec:Jetdijet}

\subsection{Jet reconstruction}

In hadronic interactions, jets originate from the fragmentation and hadronization of quarks or gluons following a hard scattering. This non-perturbative process transforms the initial high-energy partons into a dense ensemble of particles, primarily hadrons, providing a direct link to the dynamics of the underlying quarks and gluons involved in the interaction.\par

This measurement employs standardized procedures for jet reconstruction, consistent with those used in previous STAR analyses~\cite{ref:STAR_2009_Endcap_ALL}. This standardization ensures reliability and comparability across different datasets and analyses. The anti-$k_{\mathrm{T}}$ algorithm, a widely used method in jet physics, is integrated into the FastJet 3.0.6 package~\cite{ref:FastJet} for this purpose. This algorithm is preferred for its resilience to soft radiation. A resolution parameter $R$ of 0.6 is chosen, balancing the need to capture most of the jet energy while minimizing the inclusion of background particles.
\par

The inputs for the jet reconstruction include charged particle tracks detected by the TPC and energy depositions in the calorimeter towers of the BEMC and EEMC. Specific criteria are applied to these inputs to ensure accuracy and reduce background noise. Tracks are required to have a transverse momentum ($p_{\mathrm{T}}$) of at least 0.2 GeV/$c$, and must exhibit a certain quality in their detection, such as having more than five hits in the TPC and at least 51\% of the maximum possible hits along their trajectory. This requirement enhances the momentum resolution and aids in eliminating split tracks. To further refine the quality of data, a distance of closest approach (DCA) to the vertex, dependent on the $p_{\mathrm{T}}$ of the tracks, is imposed. Tracks with \( p_{\mathrm{T}} \) below 0.5 GeV/\(c\) were required to have a DCA of less than 2 cm, while those with \( p_{\mathrm{T}} \) above 1.5 GeV/\(c\) required a DCA under 1 cm. For intermediate \( p_{\mathrm{T}} \) values, a linear interpolation between these two DCA limits was applied. This criterion is crucial for filtering out tracks that are not associated with the primary hard scattering event, thus reducing the contamination from pile-up effects. For calorimeter towers, an energy threshold of $E_{\mathrm{T}} \ge 0.2$ GeV is set. Additionally, to avoid double counting the energy of charged particles that hit the calorimeter towers, a correction is applied: the transverse momentum of the track times $c$ ($p_{\mathrm{T}} c$) is subtracted from $E_{\mathrm{T}}$ of the tower. In cases where the track's momentum exceeds the tower's transverse energy, the $E_{\mathrm{T}}$ of the tower is set to zero. This approach enhances the jet momentum resolution by reducing the sensitivity to fluctuations in hadronic energy deposition, as demonstrated in previous STAR analyses~\cite{ref:Adamczyk_2015_ALL}.\par

\subsection{Dijet selection}

The dijet selection process closely follows the methodology employed in previous measurements~\cite{ref:STAR_2009_Endcap_ALL}. A basic criterion for the selection of events is the positioning of the event vertex along the beam-axis, which is required to be within 90 cm of the STAR detector's center. A dijet is chosen by selecting the two jets with the highest $p_{\mathrm{T}}$ within the pseudorapidity range $\eta_{\rm jet}$ (relative to the event vertex) between $-0.8$ and $1.8$, and a `detector pseudorapidity' $\eta_{\rm Det}$ (relative to the center of STAR) between $-0.7 \le \eta_{\rm Det} \le 1.7$. These requirements are crucial for ensuring the selected jets are within the geometric acceptance of the detector's tracking and calorimetry systems.\par

For the purpose of analysis and discussion, the jets are classified based on their pseudorapidity values. Jets within the pseudorapidity range of $-0.8 \le \eta_{\rm jet} \le 0.8$ are termed ``Barrel jets", as they fall within the barrel region of the detector. Meanwhile, jets with pseudorapidities ranging from $0.8 \le \eta_{\rm jet} \le 1.8$ are labeled as ``Endcap jets", indicating their detection in the endcap region of the detector. Only dijets that include at least one ``Endcap jet’’ are kept for the current analysis. The results for dijets with both jets falling within $-0.8 \le \eta_{\rm jet} \le 0.8$ have been published in~\cite{ref:STAR_2015results_ALL}. Thus, the lower limit of $\eta_{\rm jet} = 0.8$ for ``Endcap jets" avoids publishing overlapping datasets.\par

Jets resulting from a partonic hard-scattering event are expected to exhibit a roughly back-to-back alignment in azimuth ($\phi$). To enhance the selection of events likely originating from a $2 \rightarrow 2$ hard scattering process, an opening angle cut is applied. This cut requires the azimuthal angle between the two highest $p_{\mathrm{T}}$ jets in the event to exceed $120^\circ$. To facilitate comparisons with theoretical predictions, an asymmetric condition on the transverse momentum of the jets is imposed. Specifically, the leading jet (the jet with the highest $p_{\mathrm{T}}$ in the dijet pair) must have a $p_{\mathrm{T}}$ of at least 8.0~GeV/$c$, while the subleading jet must have a $p_{\mathrm{T}}$ of at least 6.0~GeV/$c$~\cite{Frixione:1997ks}.\par

Events that contain a track with $p_{\mathrm{T}}$ exceeding 30~GeV/$c$ are excluded if the dijets show highly imbalanced transverse momenta (with ratios greater than 3/2 or less than 2/3). This condition addresses potential inaccuracies due to the finite resolution in the track curvature calculation, which can occasionally result in a significant overestimation of a track's $p_{\mathrm{T}}$. To minimize potential biases introduced by the non-overlapping jet patch along the azimuthal direction, especially near the trigger thresholds, it is required at least one of the jets must geometrically match with the triggered jet patch. Additionally, the geometrically matched jets must exceed a minimum $p_{\mathrm{T}}$ threshold of 6.0 GeV/$c$ for JP1-triggered events or 8.4 GeV/$c$ for JP2-triggered events.\par

Furthermore, to minimize the contamination from non-collision sources, such as cosmic rays and beam-gas interactions, dijet candidates with both jets without tracks are excluded.\par


\subsection{Underlying event corrections}\label{subsec:UE_corrs}

The presence of a diffuse background in hadronic jet production presents a significant challenge in accurately reconstructing the properties of jets. This background arises from the multi-parton interactions and soft interactions between scattered partons and beam remnants, collectively referred to as the underlying event. While these particles do not participate in the primary hard scattering event that produces the jets of interest, their contributions to the energy and transverse momentum within the jet reconstruction area can significantly affect the measurement of jet properties. During 2015 operations, a higher instantaneous luminosity compared to 2009 led to an increase of the pile-up events. Additionally, the presence of the Heavy Flavor Tracker (HFT)~\cite{ref:STAR_HFT} at the center of the detector introduced more material, thereby increasing the background levels.\par 

Because the underlying event particles are generally evenly distributed across the $\eta$-$\phi$ space, it is reasonable to assume that the UE energy density is uniform.
The STAR detector exhibits good uniformity in the azimuthal angle ($\phi$) but not in pseudorapidity ($\eta$). The non-uniformity in $\eta$ is particularly noticeable in the Endcap region, where there is a service gap between BEMC and EEMC ($\Delta \eta \sim 0.08$). Also, the tracking efficiency of the TPC drops rapidly in more forward regions due to the fewer hits at intermediate pseudorapidities. To effectively address the challenges posed by variations in detector performance and the increased background, the off-axis cone method, which has been widely used in STAR jet and dijet measurements~\cite{ref:STAR_2009_Endcap_ALL,ref:STAR_2012_ALL,ref:STAR_2015results_ALL,ref:STAR_2013_ALL,ref:STAR_2012_2015_Collins}, is employed.\par

In this method, for each reconstructed jet, two off-axis cones are defined, each with a radius equivalent to the jet resolution parameter ($R = 0.6$). These cones are centered at the same pseudorapidity $\eta$ as the jet but are offset in the azimuthal direction $\phi$ by $\pm \pi/2$, positioning them away from the direct influence of the jet itself. The transverse momentum density $\rho_{\mathrm{T},UE}$ within these cones is determined by summing the $p_{\mathrm{T}}$ of all particles contained within the two cones and dividing by their total area $2\pi R^2$. This density serves as a measure of the UE's contribution to the transverse momentum observed in the vicinity of the jet. Similarly, the mass density $\rho_{\mathrm{m},cone}$ of the off-axis cones is calculated. This involves computing the invariant mass of the vector sum of all particles inside each cone, and then dividing this mass by the combined area of the two cones. This step provides an estimate of the mass contribution of the UE to the jet measurement.\par

\begin{figure}[!hbt]
    \includegraphics[width=1.0\columnwidth]{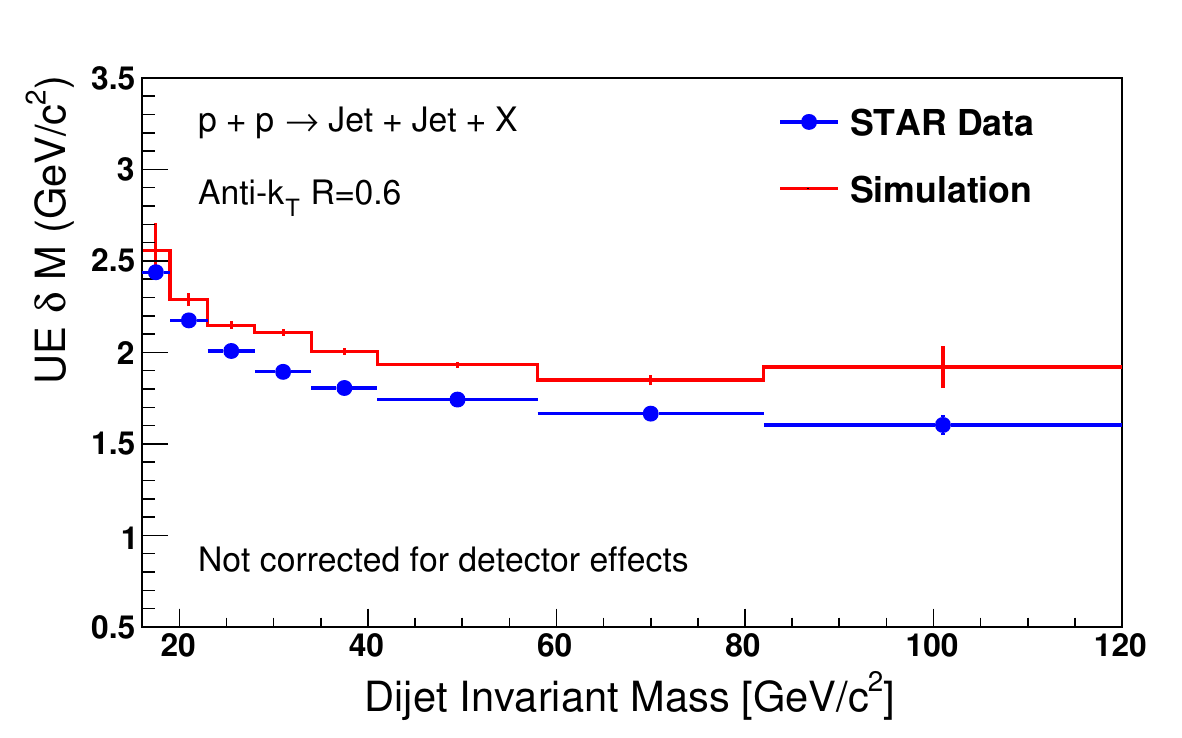}
    \caption{Data/simulation comparisons of the mean UE correction $\delta M$ as a function of underlying event corrected 
detector-level dijet invariant mass. The points represent the data and the histogram is the simulation.}
    \label{fig:dijet_UE_Mass}
\end{figure}

Dijet analyses are particularly sensitive to both the directional and mass properties of jets. To accurately reflect the impact of the UE on these properties, a four-momentum correction is applied to each jet, a process facilitated by the four-vector subtraction method developed by the FastJet group~\cite{ref:FastJet}. The equation used is:
\begin{equation}
P^{\mu}_{jet,corr} = P^{\mu}_{jet} - [\rho A_{jet}^{x}, \rho A_{jet}^{y}, (\rho+\rho_{m}) A_{jet}^{z}, (\rho+\rho_{m}) A_{jet}^{E}]
\end{equation}
where $P^{\mu}_{jet}$ and $P^{\mu}_{jet,corr}$ are the jet 4-momentum vector before and after the underlying event subtraction. $\rho$ and $\rho_m$ are the average underlying event transverse momentum and mass densities respectively, and $A_{\mu}$ is the 4-momentum vector area, as calculated by the Fastjet package~\cite{ref:FastJet} using the ghost particle technique~\cite{ref:Ghost_Particle}.\par

This correction ensures that the measured properties of dijets, such as their invariant mass, are representative of the jets' characteristics in the absence of UE contributions. Both data and simulation have UE subtraction at the detector level for dijets, and simulation also has the same procedure at the particle level. The systematic uncertainty associated with the UE correction is evaluated by comparing the corrections applied to the dijet invariant mass ($\delta M$) between the data and the simulations, as illustrated in Fig.~\ref{fig:dijet_UE_Mass}. This comparison helps to quantify the potential discrepancies in the underlying event modeling and its impact on the analysis.\par

\subsection{Comparison to simulation}\label{sec:Compare_to_sim}

\begin{figure*}
    \includegraphics[width=2.0\columnwidth]{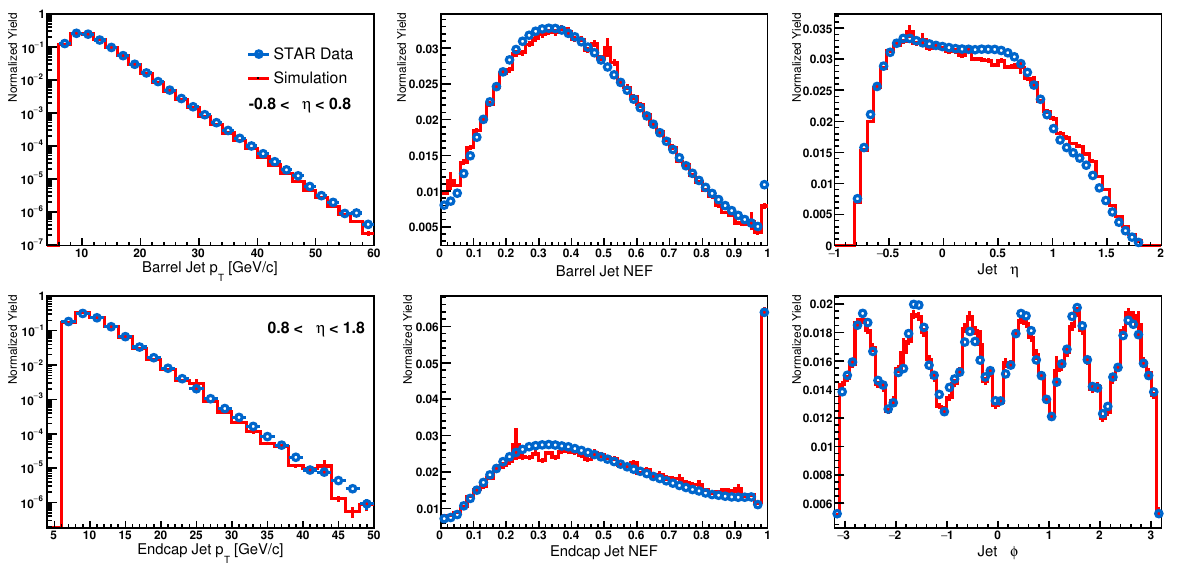}
    \caption{Data/simulation comparisons of the normalized detector level jet yields from dijet pairs as functions of detector jet transverse momentum (left) or jet neutral energy fraction (NEF, middle), for the Barrel (upper panels) and Endcap (lower panels) jets, respectively. The right two panels are plotted as functions of Barrel+Endcap jet pseudorapidity (upper right) and jet azimuthal angle (lower right). The open circles represent the data, and the histograms are the simulation.}
    \label{fig:figure1}
\end{figure*}

\begin{figure}[!hbt]
    \includegraphics[width=1.0\columnwidth]{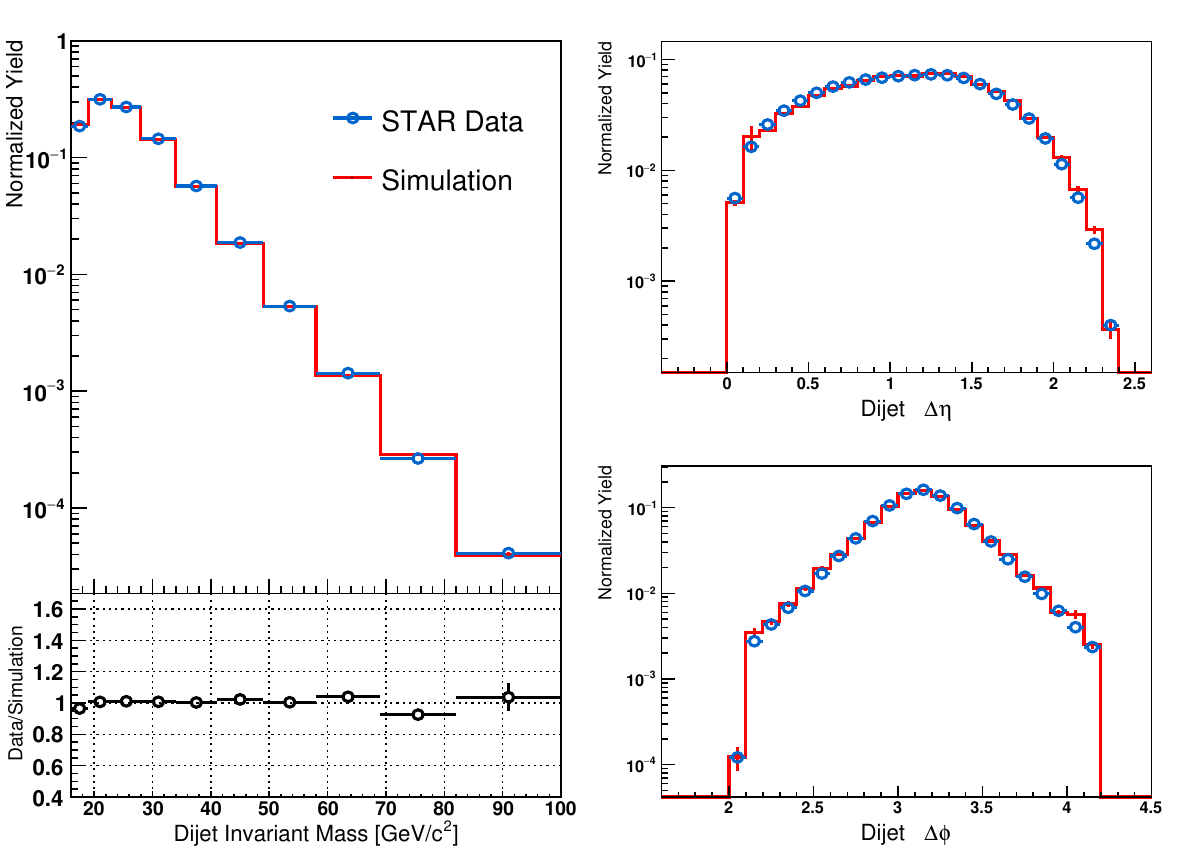}
    \caption{Data/simulation comparisons of detector level dijet yields as a function of invariant mass (left), the pseudorapidity gap (upper right) and azimuthal opening angle (lower right) between the jets for Barrel-Endcap dijets topology. The points are the data and the histograms are the simulation.}
    \label{fig:figure2}
\end{figure}

In the simulation samples, jets are reconstructed at the detector-level using the same algorithms as in the real data. The left two panels in Fig.~\ref{fig:figure1} compare the distribution of normalized jet yields as a function of detector jet $p_{\mathrm{T}}$ in both data and simulation for Barrel jets (top panel) and Endcap jets (bottom panel). The middle two panels in Fig.~\ref{fig:figure1} present a comparison between data and simulation for the observed neutral energy fraction distributions, separately for Barrel and Endcap jets. The right two panels in Fig.~\ref{fig:figure1} show the comparisons between data and simulation for jet pseudorapidity and jet azimuthal angle distributions. The simulation provides a reasonable description of all these observables.\par

Figure~\ref{fig:figure2} presents several detailed comparisons between data and simulated dijet distributions. The left panel shows the invariant mass spectrum of dijets, specifically within a topology where one jet originates from the Barrel region and its counterpart from the Endcap. The right two panels provide a detailed look at the geometrical and kinematic relationships between these jets constituting the dijet, highlighting the distributions of pseudorapidity differences ($\Delta\eta$) and azimuthal angle differences ($\Delta\phi$, representing the opening angle). Good agreement is also observed for these distributions.\par

Jets are also reconstructed in simulation at the particle and parton levels using the anti-$k_{T}$ algorithm with a resolution parameter of $R = 0.6$. As mentioned earlier, particle-level dijets consist of all stable final-state particles in Perugia 2012 tune~\cite{ref:PerugiaTunes}, while parton-level dijets are reconstructed from all hard-scattered partons produced in the collision, including those from initial- and final-state radiation, but not those from the underlying event or beam remnants. The same off-axis cone UE correction procedure used for detector jets is applied to the particle jets. It's important to note that the selection of dijets at the particle or parton levels from the comprehensive, unbiased \textsc{Pythia} sample does not necessitate considerations for the detector's performance. As such, specific cuts typically applied in detector-level analyses, such as the neutral fraction cuts, are not relevant and thus not implemented at these levels.\par

Detector-level jets, on the other hand, are subject to the physical limitations and efficiencies of the actual detection equipment, making it essential to understand how these practical constraints affect jet measurements. Establishing a correlation between the jets reconstructed at the particle or parton levels and those observed at the detector level is vital for accurately determining corrections and assessing systematic uncertainties. In practice, this correlation process involves first identifying a dijet at the detector level. Subsequently, particle and parton level dijets are associated with this detector-level dijet if both of the jets within the pair fall within a defined matching criterion, specifically a radius $\Delta R = \sqrt{\Delta\eta^2 + \Delta\phi^2} < 0.5$. If more than one parton- or particle-level jet matches a given detector jet, the one closest in $\eta$–$\phi$ space is selected. Dijet pairs are considered matched if each jet in the pair satisfies the inclusive jet matching criterion. The detector–parton dijet matching fraction is approximately 98\% for dijet invariant masses in the range $M = 16 – 19~\mathrm{GeV}/c^{2}$, and quickly approaches 100\% for $M > 19~\mathrm{GeV}/c^{2}$. The requirement to reconstruct two nearly back-to-back jets significantly improves the matching efficiency compared to the inclusive jet case. This greatly reduces the reconstruction of fake jets or jets with poorly reconstructed axes due to contributions from the underlying event or background. Any residual dilution from fake jets is accounted for in the trigger and reconstruction bias correction (see Sec.~\ref{sec:trigbias} for details).\par

\section{Tracking inefficiency and correction methods}\label{sec:Methods}

\begin{figure}[!hbt]
    \includegraphics[width=1.0\columnwidth]{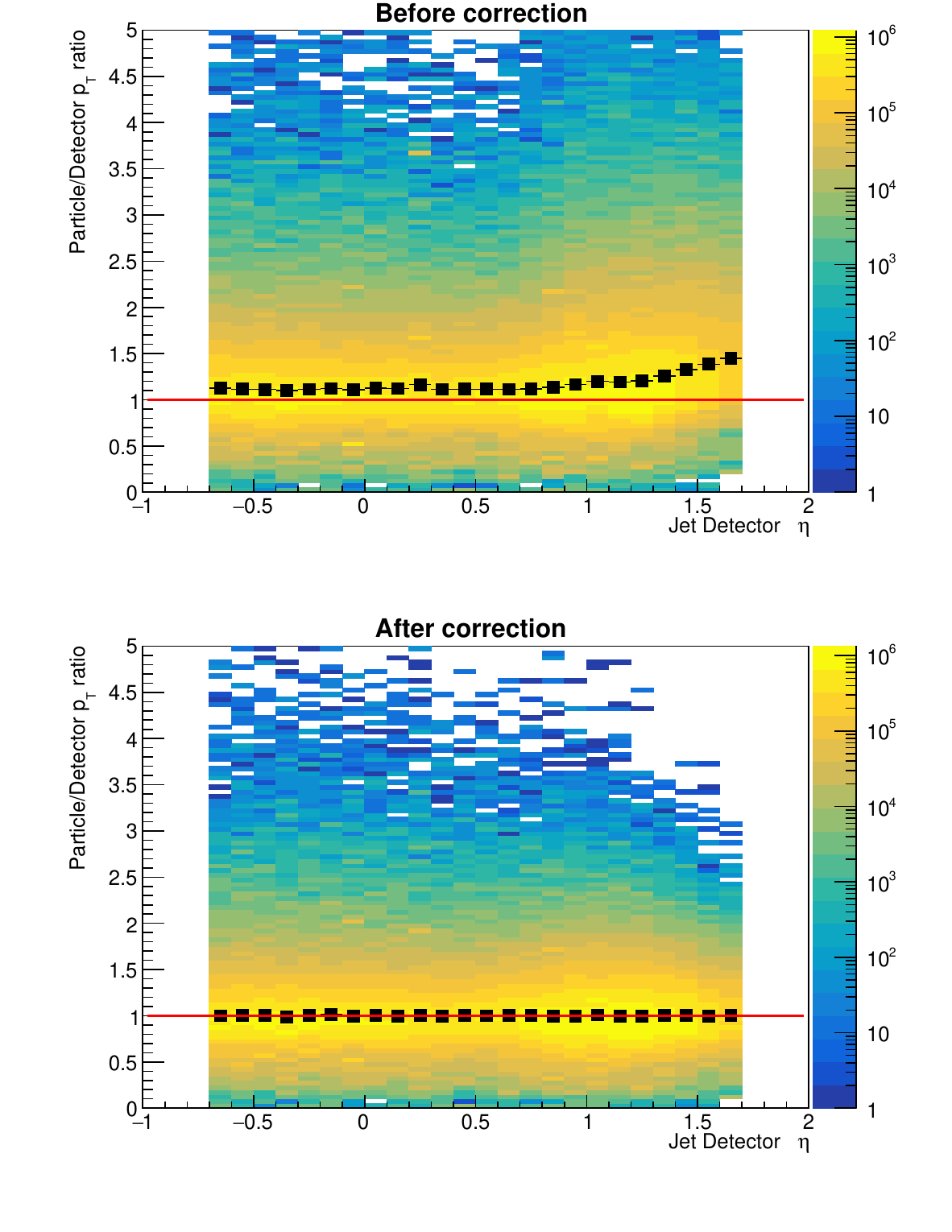}
    \caption{Particle-level $p_{\mathrm{T,jet}}$ divided by detector-level $p_{\mathrm{T,jet}}$ as a function of detector $\eta$ before (upper plot) and after (lower plot) a $p_{\mathrm{T}}$ correction was made. The correction is determined using machine-learning techniques. The black points are the average values.}
    \label{fig:figure_pt_correction}
\end{figure}

\begin{figure}[!hbt]
    \includegraphics[width=1.0\columnwidth]{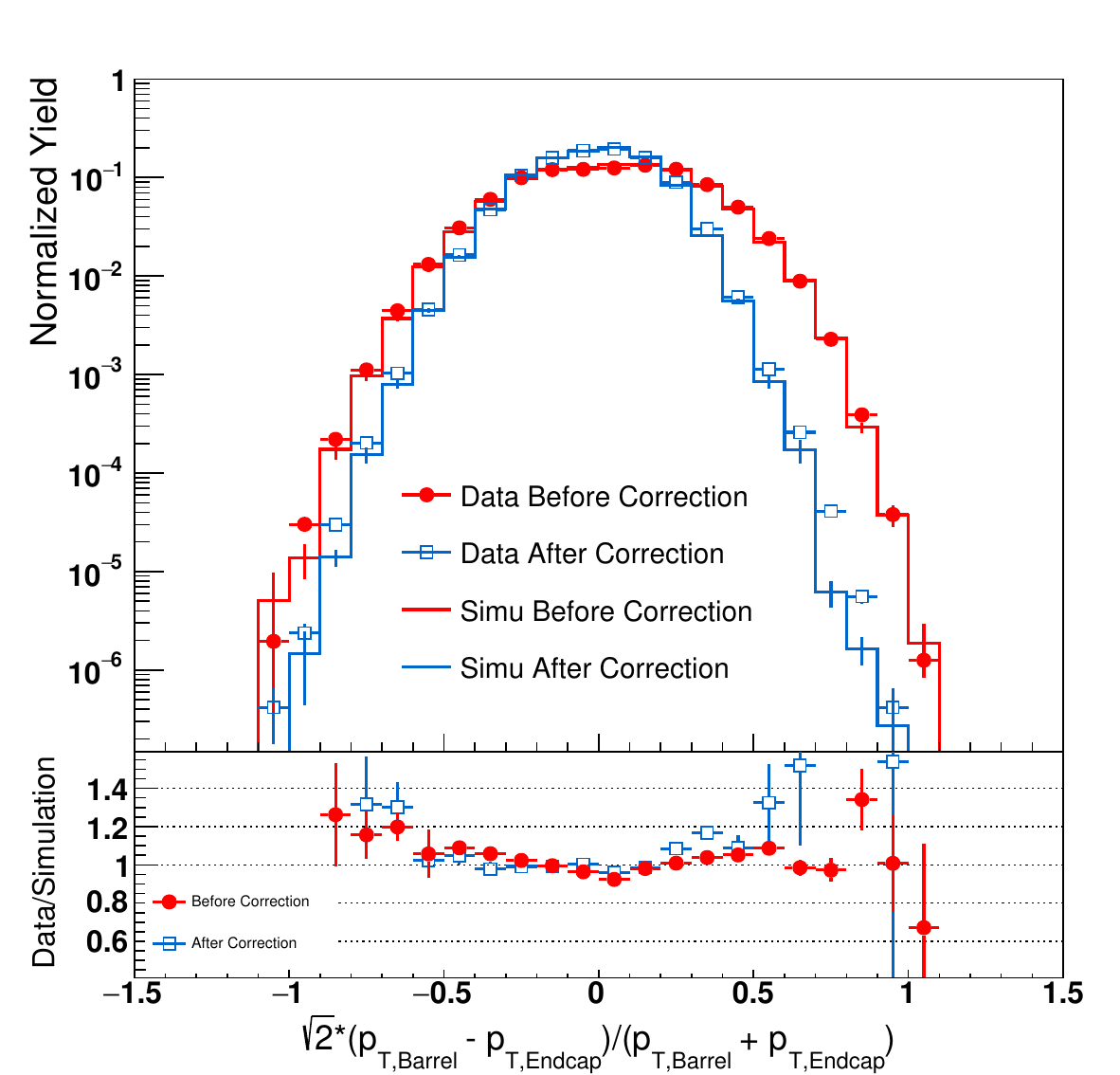}
    \caption{The dijet $p_{\mathrm{T}}$ imbalance distribution for the Barrel-Endcap dijet topology before (red) and after (blue) $p_{\mathrm{T}}$ corrections were made. The points represent the data (solid circle for before and open square for after the correction), and the histograms are the simulation.}
    \label{fig:figure_pt_imbalance}
\end{figure}

The primary challenge in this measurement arises from the reduced tracking efficiency in the EEMC region. While the TPC maintains its efficiency within the nominal pseudorapidity range of $|\eta| \le 1.3$, there is a notable decrease in the forward direction, including a significant portion of the EEMC's coverage. This diminished tracking efficiency leads to a tendency for jets in the Endcap to be reconstructed with lower $p_{\mathrm{T}}$ on average. The implications of this effect are clearly demonstrated in the simulation, as shown in the upper panel of Fig.~\ref{fig:figure_pt_correction}. This plot presents the ratio of particle-level jet $p_{\mathrm{T}}$ to detector-level $p_{\mathrm{T}}$ across various detector $\eta$ values. The measured jet $p_{\mathrm{T}}$ is systematically lower than its particle level jet $p_{\mathrm{T}}$. Thus, it complicates the accurate extraction of initial state parton momenta and favors the selection of jets with a higher proportion of neutral energy, skewing the sample towards jets less representative of the charged particle energy distribution. Additionally, the reconstruction process biases the jet mass calculation. In the jet reconstruction algorithm, charged particle tracks are assumed to have the mass of pions, whereas calorimeter towers are treated as massless, leading to an underestimation of the detector-level jet invariant mass compared to its true value.\par

To address these challenges and improve the accuracy of jet $p_{\mathrm{T}}$ and mass estimations, a machine learning approach was adopted. A Multilayer Perceptron (MLP), provided by the Toolkit for Multivariate Data Analysis (TMVA)~\cite{ref:tmva}, was utilized as a regression method. This technique is adept at correcting both the jet $p_{\mathrm{T}}$ and mass as determined by the jet finding algorithm, thereby enhancing the accuracy of the measurement and mitigating the biases and limitations introduced by the detector's tracking inefficiencies and reconstruction assumptions.\par

In the regression analysis, all simulated events containing Endcap jets were utilized, which were further separated into training and testing subsamples. The primary variables selected for the $p_{\mathrm{T}}$ correction process were the measured jet $p_{\mathrm{T}}$ and the detector pseudorapidity. The preference for detector pseudorapidity over the jet's own $\eta_{\rm jet}$ value was strategic, chosen for its direct relevance to the detector's geometry that significantly influences tracking efficiency. Additionally, the jet's neutral energy fraction was incorporated as an input variable, providing critical insight into the biases that arise from the diminished tracking efficiency in the Endcap region. Moreover, considering that jets in dijet events are expected to have approximately equal transverse momenta, the $p_{\mathrm{T}}$ of the away-side jet located in the Barrel, which is reconstructed with greater precision due to better tracking efficiency, was also included as an input for the Endcap jet's $p_{\mathrm{T}}$ correction. As mentioned earlier, particle-level to detector-level jet association is accomplished by iterating over all particle-level jets and selecting the one closest in $\eta$-$\phi$ space, with a geometric matching condition of a distance less than 0.5. The target for the regression analysis (the value that the $p_{\mathrm{T}}$ correction aims to achieve) is the $p_{\mathrm{T}}$ of the particle-level jet. This method integrates the detailed jet characteristics and detector-specific parameters to refine the accuracy of jet $p_{\mathrm{T}}$ measurements, thereby enhancing the overall precision of the correction.\par

The neural network was trained with its parameters optimized to enhance the performance. The effect of this machine learning correction is illustrated in the lower panel of Fig.~\ref{fig:figure_pt_correction}, which presents a comparison between the neural network's output and the target values (the ratio of particle-level jet $p_{\mathrm{T}}$ to the corrected detector-level jet $p_{\mathrm{T}}$). This ratio exhibits a remarkable consistency across varying detector pseudorapidities after correction, as shown by the data points with uncertainty bars (which are smaller than the data points). These points denote the average within each pseudorapidity bin. Additionally, the vertical spread in the distribution is reduced, resulting in an average improvement of approximately $34\%$ in the resolution of jet transverse momentum.\par

Similar jet $p_{\mathrm{T}}$ corrections were also applied to the Barrel region. Although the jet transverse momentum is generally reconstructed more accurately in the Barrel than in the Endcap, a systematic underestimation of the measured $p_{\mathrm{T}}$ relative to its true value was observed, due to the limitations in detector performance. The correction method employed for Barrel jets mirrors the one described above, with the exception that the correlated jet $p_{\mathrm{T}}$ from the opposing (Endcap) jet was not incorporated as an input in the correction process for Barrel jets. This ensures that both Barrel and Endcap jets undergo precise $p_{\mathrm{T}}$ correction, significantly improving the accuracy of jet measurement across the detector.\par

Figure~\ref{fig:figure_pt_imbalance} presents the distribution of the dijet $p_{\mathrm{T}}$ imbalance, which is defined as the difference in the transverse momentum magnitudes between the two jets in Barrel-Endcap dijet events. Initially, before any corrections (depicted by the red curve), there is a noticeable tendency for the reconstructed jet $p_{\mathrm{T}}$ in the Barrel to surpass that of its counterpart in the Endcap. This discrepancy leads to a systematic bias towards positive values in the $p_{\mathrm{T}}$ imbalance distribution. However, following the application of corrections (illustrated by the blue curve), this systematic difference is significantly reduced, and the distribution's spread becomes noticeably narrower in both the data and the simulations.\par

Although jet mass is generally small relative to its transverse momentum at RHIC kinematics, it is a crucial jet property for determining the dijet invariant mass. To refine the accuracy of jet mass measurements, machine learning techniques were similarly employed, following the approaches previously outlined for $p_{\mathrm{T}}$ corrections. The inputs for the artificial neural network included the initially calculated jet mass, track and tower multiplicities, jet transverse momentum, the fraction of neutral energy, and detector pseudorapidity. The decision to incorporate jet $p_{\mathrm{T}}$, neutral energy fraction, and detector pseudorapidity as inputs was motivated by the observed influence of decreasing tracking efficiency on the accuracy of jet mass estimation. Similarly, the goal of this correction process was to match the detector-level jet mass with its particle-level counterpart from the simulation, thereby enhancing the precision of jet property measurements and contributing to a more accurate representation of dijet dynamics. For jets in dijet events where both jets are located within the Endcap region, the corrections are made with the same variables as those employed for Barrel jets in Barrel-Endcap dijet configurations. Similar improvements are also seen after the corrections.\par

\section{The spin asymmetry $A_{LL}$}\label{sec:asymmetry}

At RHIC, the spin observable that most directly probes the helicity distribution of gluons within the proton, denoted as $\Delta g(x)$, is the longitudinal double-spin asymmetry, $A_{LL}$. It is defined as the ratio of the helicity-dependent cross sections:
\begin{equation}
A_{LL} \equiv \frac{\sigma_{++} - \sigma_{+-}}{\sigma_{++} + \sigma_{+-}}
\label{equ:ALL_cross}
\end{equation}
where $\sigma_{++}$ and $\sigma_{+-}$ represent the differential production cross sections corresponding to proton beams with equal and opposite helicities, respectively. Experimentally, the asymmetry is determined by sorting the measured yields based on the beam spin state and combining this information with other independent measurements. The expression for the asymmetry is given by:
\begin{equation}
A_{LL} = \frac{\sum(P_{Y}P_{B})(N^{++}-rN^{+-})}{\sum(P_{Y}P_{B})^{2}(N^{++}+rN^{+-})}
\label{equ:ALL}
\end{equation}
where the ratio of the differential cross section is further converted into the ratio of dijet yields ($N$), which is then scaled by the luminosity ratio ($r$) for the same ($++$) and opposite ($+-$) helicity beams. The summation in Eq.~\ref{equ:ALL} is over independent data runs, with run lengths ranging from 10 to 60 minutes in 2015. The beam polarizations $P_{B}$ and $P_{Y}$ are calculated luminosity-weighted and time-dependent (provided by the RHIC polarimetry group~\cite{ref:RHICPolG}) for each run. It is noteworthy that the run duration is relatively short compared to the time scale over which the beam polarizations and relative luminosities varied, {\it e.g.} the average lifetime is about 60 hours for polarization.\par

\subsection{Dijet invariant mass correction}
\begin{figure}[!hbt]
    \includegraphics[width=1.0\columnwidth]{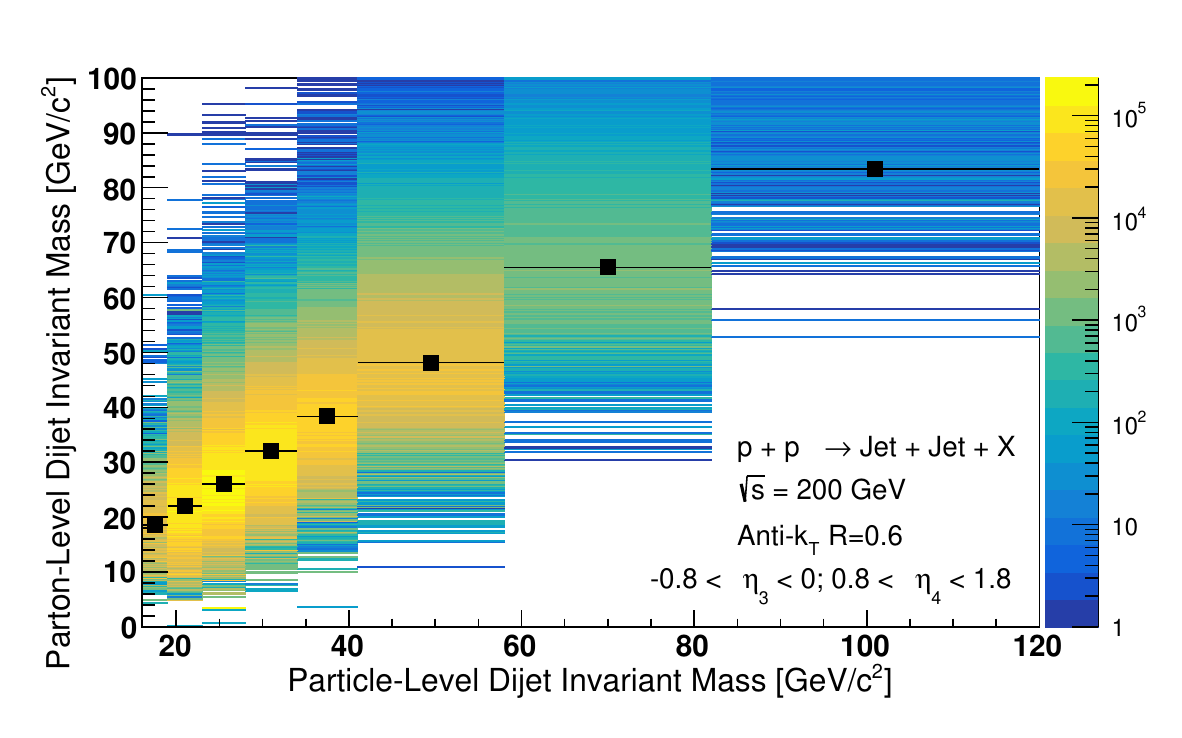}
    \caption{Dijet invariant mass correlation between detector-level and parton-level for East Barrel-Endcap dijet topology.}
    \label{fig:dijet_Mass_Corr}
\end{figure}

To facilitate the comparison between our experimental results and theoretical predictions, which are calculated at the parton level, we applied a simple mass shift to each data point to determine the parton-level dijet invariant mass. Compared to the exponentially varying cross sections, the asymmetries are small and exhibit only a weak, nearly linear dependence on the dijet invariant mass. Consequently, a bin-by-bin correction is applied to ensure that the final asymmetry values are reported at the average parton-level dijet mass for each particle-level dijet mass bin. This adjustment accounts for the difference between the parton-level and particle-level dijet invariant masses as shown in Fig.~\ref{fig:dijet_Mass_Corr}. As detailed in the previous section, the machine learning procedure corrected jets back to the particle level. Thus, this mass shift was calculated by comparing particle-level masses to their corresponding parton-level dijet masses. For each particle-level mass bin, we calculated the event-by-event difference between parton and particle-level dijet masses. The correction was then determined as the mean value of these differences, averaged across the entire simulation sample. The final data points are plotted at this average particle-level mass plus the estimated particle-to-parton mass shift.\par

\subsection{Trigger and reconstruction bias}\label{sec:trigbias}
In hadronic $pp$ interactions, quark-quark ($qq$), quark-gluon ($qg$), and gluon-gluon ($gg$) are the three dominant partonic scattering processes. Thus, the value of $A_{LL}$ extracted from the data represents an admixture of the asymmetries produced from the three partonic scattering subprocesses. However, the efficiency of the STAR jet patch trigger system might not be uniform across all these processes, which could potentially skew the fractions of each subprocess contributing to the observed data. This variation in trigger efficiency may, in turn, influence the measured $A_{LL}$ values. Furthermore, systematic deviations can occur due to the finite resolution of the detector. When combined with the rapidly falling invariant mass distribution, these resolution effects can lead to shifts in the subprocess fractions associated with specific mass values.\par

To address these challenges and accurately reflect the underlying partonic dynamics, a trigger and reconstruction bias correction was applied to the raw $A_{LL}$ measurements. This correction is designed to neutralize the impact of trigger efficiency variations and detector resolution limitations, thereby ensuring that the final $A_{LL}$ values accurately reflect the underlying partonic interactions without distortion.\par

In this method, the parameterizations of the polarized parton distribution functions are combined with \textsc{Pythia} parton kinematic variables to generate predictions of $A_{LL}$ vs. dijet mass for a particular model at both the parton and detector levels. The asymmetry is calculated by first extracting the parton scattering kinematics, including the Mandelstam variables $u$, $s$, and $t$ for the parton-parton scattering process, as well as the parton flavors involving in the $2 \rightarrow 2$ scattering. This allows for the calculation of the ratio of the polarized to unpolarized partonic cross-sections at the leading order. The result is then multiplied by the ratio of the polarized to unpolarized PDFs for the two incoming partons at their respective $x_{1}$, $x_{2}$ and $Q^{2}$ as shown in the equation:

\begin{equation}
A_{LL} = \frac{\Delta \sigma}{\sigma} \times \frac{\Delta f_{1}(x_{1},Q^{2})}{f_{1}(x_{1},Q^{2})} \frac{\Delta f_{2}(x_{2},Q^{2})}{f_{2}(x_{2},Q^{2})},
\end{equation}
Here $\Delta \sigma$ represents the polarized partonic cross-section, $\sigma$ is the unpolarized partonic cross-section, $\Delta f(x,Q^{2})$ is the polarized PDF, and $f(x,Q^{2})$ is the unpolarized PDF. Similar to previous measurements, the NNPDFPol1.1 polarized PDF set~\cite{ref:NNPDF} was used as input. The associated ensemble of publicly available replica sets provides a robust framework for estimating the uncertainty on the correction. These replicas, generated using a Monte Carlo sampling approach, represent statistically independent realizations of the PDFs, each consistent with the experimental constraints included in the global fit. The full ensemble is used to extract both central values and associated uncertainties of derived observables. The trigger and reconstruction bias correction for each mass bin was determined by evaluating the quantity
\begin{equation}
\Delta A_{LL} = A^{det}_{LL} - A^{parton}_{LL}
\end{equation}
for each of the 100 replicas in the NNPDFPol1.1 set. The value of $A_{LL}^{det}$ represents the measured longitudinal double-spin asymmetry for detector-level dijets in the simulation, while the $A_{LL}^{parton}$ denotes the $A_{LL}$ value for pure parton-level dijets. This parton-level information is generated before incorporating the GEANT model and the trigger filter. Subsequently, it is calculated at the average parton-level dijet mass corresponding to the dijet bins sampled by the detector. The correction is derived as the average of the 100 replica values for $\Delta A_{LL}$. Then the final result is obtained by subtracting this average correction from the raw $A_{LL}$ value, yielding $A_{LL}^{final} = A_{LL}^{raw} - \Delta A_{LL}^{average}$. To determine the systematic uncertainty associated with dijet $\Delta A_{LL}$, the statistical uncertainties of the detector-level NNPDF $A_{LL}$ and the square root of the variance of the 100 $\Delta A_{LL}$ values are added in quadrature. The corrections generally fall between 0.0005 and 0.0026, accounting for roughly 10\% of the measured asymmetries. Meanwhile, the corresponding systematic uncertainties vary from 0.0003 to 0.0008.\par

\subsection{Systematic uncertainty}

The systematic uncertainties in the analysis of dijet events are crucial, influencing both the calculated dijet invariant mass (``$x$-axis uncertainties") and the measured longitudinal double-spin asymmetries ($A_{LL}$, referred to as ``$y$-axis uncertainties"). The systematic uncertainties affecting $A_{LL}$ include the beam polarization, relative luminosity, effects from the underlying event, biases introduced by the trigger and jet reconstruction processes, and residual transverse polarization. On the other hand, the systematic uncertainties impacting the dijet invariant mass calculation arise from factors such as the jet energy scale, tracking efficiency, underlying event, corrections applied to jet $p_{\mathrm{T}}$ and mass, shifts in the dijet invariant mass, and the selection of the \textsc{Pythia} tune for simulations. While some of these uncertainties have been previously discussed, others warrant further clarification.\par

The uncertainty in the product of the average beam polarizations, a relevant quantity for double-spin asymmetries, was determined by the RHIC polarimetry group and was estimated to be $6.1\%$~\cite{ref:RHICPolG}. The relative luminosity and the residual transverse polarization systematic uncertainties remain consistent with those determined for the inclusive jet and mid-rapidity dijet analyses~\cite{ref:STAR_2015results_ALL}, with values of $\pm 0.0007$ and $\pm 0.000065$, respectively, and are applied to all the mass bins.\par

\subsubsection{Dijet energy scale systematic uncertainties}
The primary source of the systematic uncertainty in the reconstructed dijet mass arises from the uncertainty in the jet energy scale. The jet energy scale uncertainties consist of two parts: one from the scale and status uncertainties of the EMC towers (3.5\% for BEMC and 4.6\% for EEMC), and the other from the TPC track transverse momentum uncertainty and the uncertainty in the tower response to charged hadrons.\par

The uncertainty attributable to the tracks, including those from track reconstruction, is found to be minimal at 0.003 based on weak decays study for 2015. The uncertainty of tracks projected to the BEMC also contribute to this part, which is estimated to be about 1.4\% based on the tracking efficiency as well as the hadronic energy deposition and response of the electromagnetic towers. The total uncertainty of tracks is scaled by the charged energy fraction for each jet. Due to reduced tracking efficiency, the systematic uncertainty calculation for Endcap jets differs from that for Barrel jets. The tracking efficiency and scale factor adjustments for Endcap jets reflect these differences, resulting in a tower-track response systematic uncertainty of 2.3\%, unaffected by the neutral energy fraction.\par

To further quantify uncertainties related to tracking efficiency, we evaluated their impact by comparing the average dijet invariant mass difference between detector and parton levels. This assessment involved using the complete set of reconstructed tracks from the TPC compared to a scenario where only a partial set of tracks was employed. The partial set was created by randomly excluding a specific percentage of tracks from the full set before conducting jet reconstruction. In our analysis, the rejection fraction was set to $7\%$, which exceeds the typical STAR tracking efficiency uncertainty, to account for the limited determination provided by short tracks at $\eta > 1$. Systematic uncertainties due to the dijet invariant mass shift also account for limited simulation sample statistics, with the overall statistical uncertainty derived by summing the uncertainties from various trigger samples in quadrature, weighted according to their trigger fractions.\par

\subsubsection{\textsc{Pythia} tune systematic uncertainties}

\textsc{Pythia} parameters can be independently adjusted to accommodate various data sets. In this study, the Perugia2012 set in \textsc{Pythia} 6.4.28 \cite{ref:Pythia6} is used as the standard tune set in the simulation, with the parameter PARP(90) fine-tuned from 0.24 to 0.213. To assess the impact of \textsc{Pythia} parameter adjustments on our results and to quantify the systematic uncertainty arising from the choice of tune, we explore a range of seven alternative tunes within the Perugia2012 set, namely 371, 372, 374, 376, 377, 378, and 383. Tunes 371 and 372 use different renormalization scales to effectively change the initial- and final-state radiation. Tune 374 reduces the color reconnection strength relative to the default, and tunes 376 and 377 adjust the fragmentation parameters to produce more longitudinal and more transverse energy distributions, respectively. In addition, tune 378 substitutes the default CTEQ6L1 PDF with the MSTW 2008 LO set, and tune 383 employs an alternative Innsbruck hadronization parameter set to further probe uncertainties in the hadronization process~\cite{ref:PerugiaTunes}.\par

For each selected \textsc{Pythia} tune, we calculate the shifts in the dijet invariant mass between the particle-level and parton-level. For tunes 371 and 372, as well as 376 and 377, which modify the same parameters but in opposing directions, we derive the uncertainty by taking half of the absolute difference between each pair's results. The final systematic uncertainty is then determined by summing these differences in quadrature across all considered tunes, and it is about 0.4 $\mathrm{GeV}/c^{2}$ on average for most data points.\par

\subsubsection{Systematic uncertainties on machine learning correction}

Corrections to jet transverse momentum and invariant mass in this analysis were conducted using the MLP, a machine learning technique known for its adaptability to complex nonlinear data relationships. Given the MLP's sensitivity to its configuration, such as the number of layers and nodes, we considered the potential influences these parameters might have on the accuracy of our corrections. To estimate the systematic uncertainties associated with the $p_{\mathrm{T}}$ and mass corrections, we compare outputs derived from slightly modified input parameters and network configurations, as well as from alternative machine learning methods, with the differences among these outputs being added in quadrature.

For the MLP, we systematically varied factors such as the training sample size, the number of layers, and the number of nodes to estimate their impact on the jet $p_{\mathrm{T}}$ and mass corrections. Additionally, to evaluate the sensitivity of our results to the choice of machine learning algorithm, we employed alternative methods, namely the Linear Discriminant (LD) and K-Nearest Neighbors (KNN) algorithms, both from TMVA. For each event, corrections to jet $p_{\mathrm{T}}$ and mass were made using these diverse configurations and methods, following which we calculated the differences in the dijet invariant masses between these modified samples and the default MLP settings. The deviations arising from the various machine learning methods and different MLP configurations were then summed in quadrature to provide the final estimation of the systematic uncertainty associated with the jet $p_{\mathrm{T}}$ and mass corrections. For most data points, it is about 0.15~$\mathrm{GeV}/c^{2}$ on average.\par

\subsubsection{Underlying event systematic uncertainties}

The presence of the underlying event in hadronic interactions results in a systematic increase in the energy scale of the jets composed solely from showers arsing from only the hard scattering. Consequently, this phenomenon introduces distortions to the measured dijet invariant mass. As already discussed in Sec.~\ref{subsec:UE_corrs}, the systematic uncertainty associated with the UE correction is evaluated by comparing the corrections applied to the dijet invariant mass between the data and the simulations.\par

Furthermore, if the underlying event exhibits a longitudinal double-spin dependence, it could induce a spin-dependent energy to the triggered jet patches, which could introduce a spin-dependence in the trigger efficiency that is not associated with the hard scattered partons. To assess the potential impact of this factor, an investigation into the spin dependence of the measured $\delta M_{helicity}$ (change in invariant mass for different helicity states) values was conducted. We did not observe any significant effect attributable to the underlying event's spin dependence. The established upper limits for the potential distortion were found to be less than 0.2\% for Barrel-Endcap dijets and less than 0.4\% for Endcap-Endcap dijets. These constraints were subsequently utilized to estimate the alterations in the dijet cross-section that could be ascribed to the UE. These estimations were then assigned as the corresponding systematic uncertainties in the analysis.\par

\begin{figure*}
  \centering
  \begin{minipage}[h]{0.49\textwidth}
    \includegraphics[width=1.0\columnwidth]{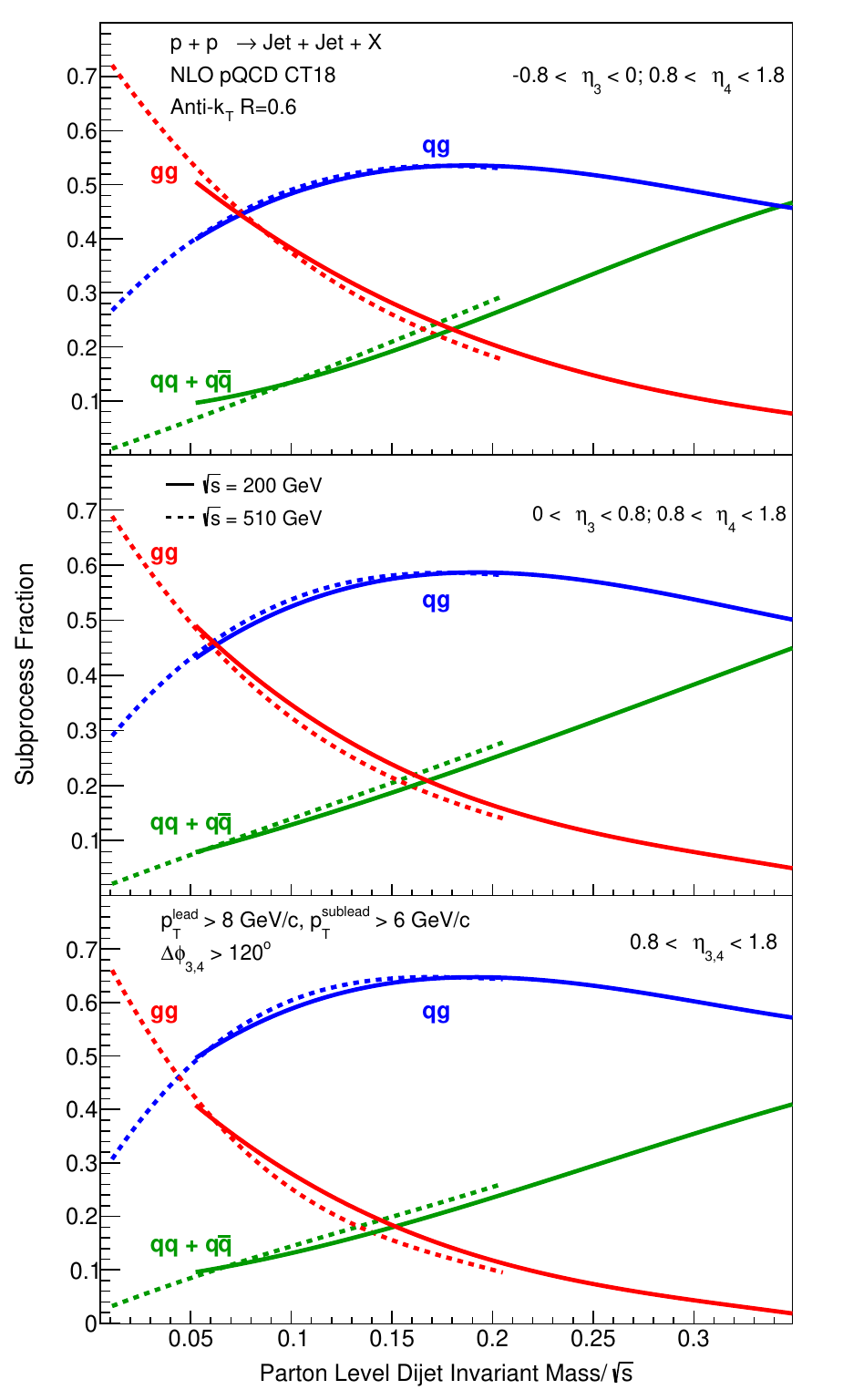}
    \caption{Fractions of the next-to-leading-order cross section for dijet production arising from quark-quark, quark-gluon, and gluon-gluon scattering in $pp$ collisions at 200 and 510 GeV, as a function of dijet invariant mass over the collision energies.}
    \label{fig:Dijet_subprocess_nloct18}
  \end{minipage}
  \hfill
  \begin{minipage}[h]{0.49\textwidth}
    \includegraphics[width=\textwidth]{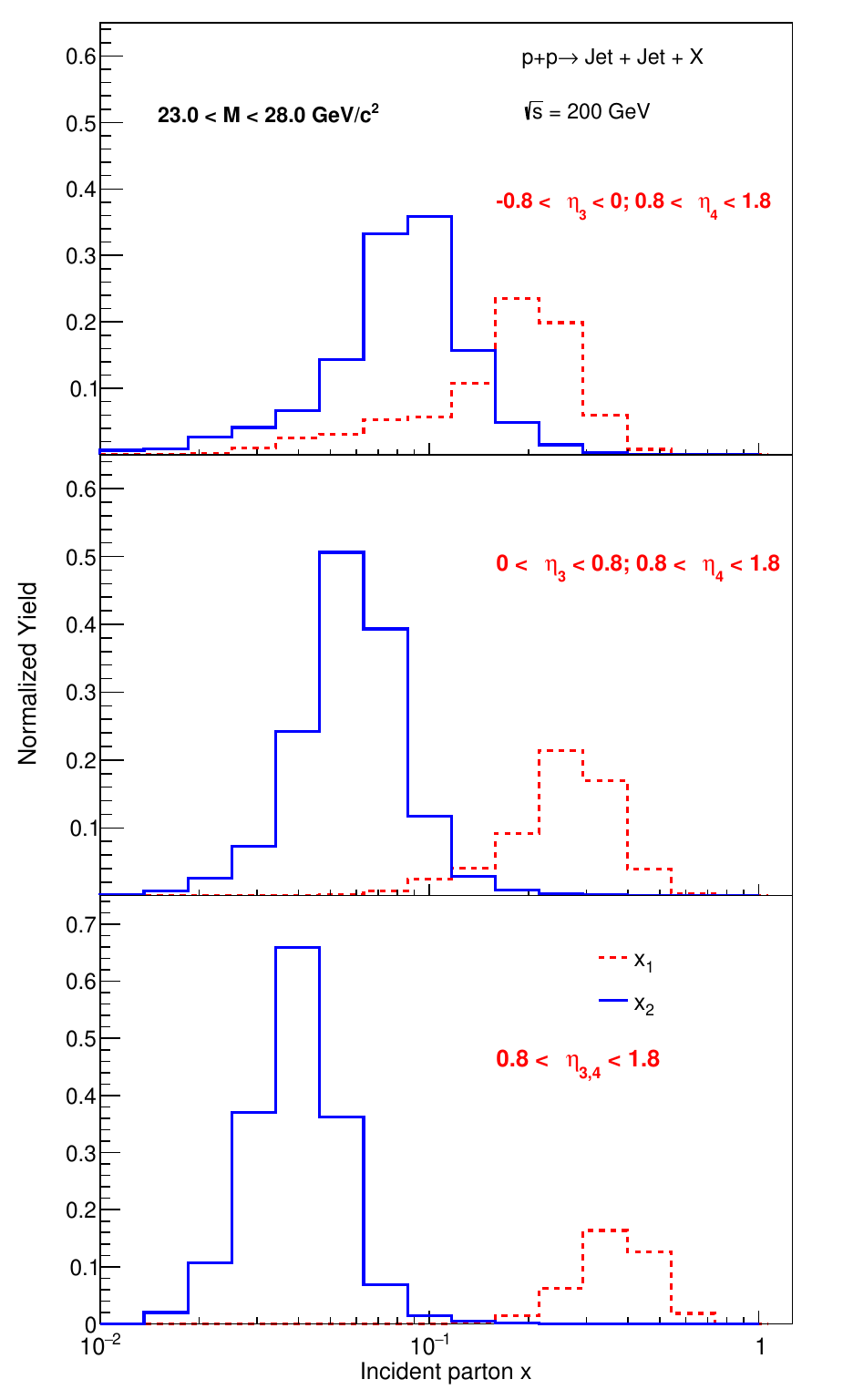}
    \caption{The distributions of the parton $x_1$ and $x_2$, which have been weighted by the partonic $\hat{a}_{LL}$, from \textsc{Pythia} detector level simulations at $\sqrt{s}$ = 200 GeV, for different jet pseudorapidity ranges. $x_1$ is associated with the beam heading towards the EEMC.}
    \label{fig:Dijet_Kinematics}
    \end{minipage}
\end{figure*}

\section{Final results}\label{sec:Results}

\begin{figure}[!hbt]
    \includegraphics[width=0.49\textwidth]{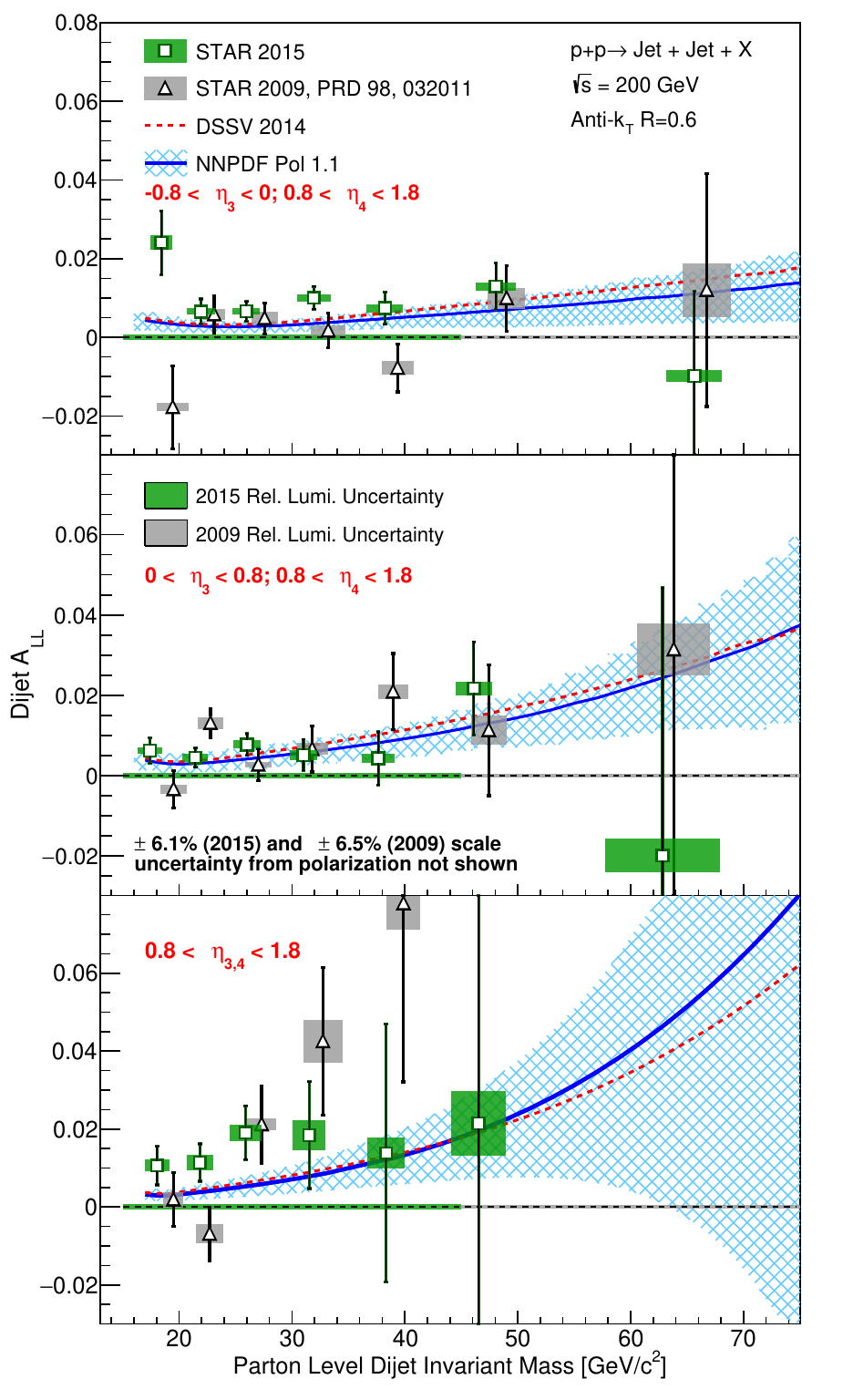}
    \caption{$A_{LL}$ as a function of parton-level invariant mass for dijets with the East Barrel-Endcap (top), West Barrel-Endcap (middle) and Endcap-Endcap (bottom) event topologies. The squares present the new results from 2015, while the triangles are the previous results with 1~GeV offset on the horizontal axis~\cite{ref:STAR_2009_Endcap_ALL}. The error bars show the statistical uncertainties. The boxes show the systematic uncertainties, with the heights giving the systematic uncertainties in $A_{LL}$ and the widths giving the parton-level dijet mass uncertainties. The theoretical expectations are based on global analysis from NNPDF Pol 1.1~\cite{ref:NNPDF} and DSSV2014~\cite{ref:DSSV2014}, with a renormalization scale of $\mu_R = p_{\text{T,jet}}$.}
    \label{fig:ALL_DiffTopo}
\end{figure}

\begin{figure}[!hbt]
    \includegraphics[width=0.49\textwidth]{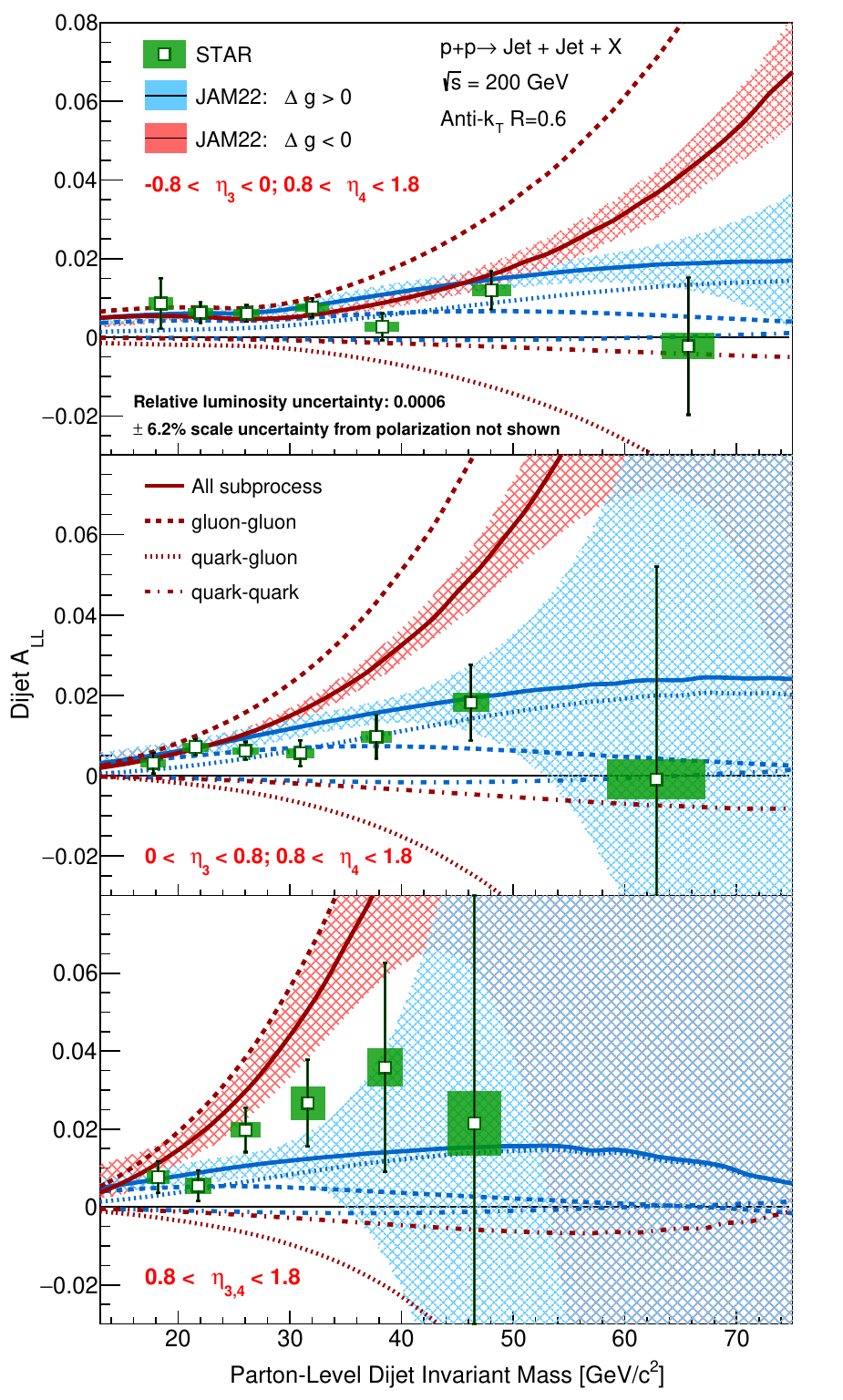}
    \caption{$A_{LL}$ as a function of parton-level invariant mass for dijets with the East Barrel-Endcap (top), West Barrel-Endcap (middle) and Endcap-Endcap (bottom) event topologies. The STAR results presented here are the combination of 2009 and 2015 measurements. The error bars show the statistical uncertainties. The boxes show the systematic uncertainties, with the heights giving the systematic uncertainties in $A_{LL}$ and the widths giving the parton-level dijet mass uncertainties. The theoretical expectations are based on global analysis from JAM22~\cite{ref:jam_negative_gluon}, with a renormalization scale of $\mu_R = p_{\text{T,jet}}$.}
    \label{fig:ALL_withJAM22}
\end{figure}

The final results are categorized into three distinct dijet topologies: dijets in which one jet is detected in the east half of the Barrel EMC ($-0.8 < \eta_{\rm jet} < 0.0$) or in the west half of the Barrel EMC ($0.0 < \eta_{\rm jet} < 0.8$), while the other is in the Endcap ($0.8 < \eta_{\rm jet} < 1.8$), and events in which both jets fall in the Endcap. As shown in Fig.~\ref{fig:Dijet_subprocess_nloct18}, RHIC dijet data exhibit direct sensitivity to gluon polarization. This is attributed to the dominance of gluon-gluon and quark-gluon scattering processes in hard scattering events at RHIC energies. Notably, the polarized $pp$ collisions at 200 GeV primarily involve quark-gluon interactions.\par

The different event topologies probe distinct ranges of the momentum fractions, $x_1$ and $x_2$, carried by the partons participating in the hard scattering, where $x_1$ is associated with the beam heading towards the EEMC. The distributions of $x_1$ and $x_2$ obtained from simulation for the three topologies are illustrated in Fig.\ref{fig:Dijet_Kinematics}. These distributions are weighted by the partonic longitudinal double spin asymmetry $\hat{a}_{LL}$~\cite{Craigie:1983qjl} appropriate for each subprocess, thereby highlighting the regions that are sensitive to gluon polarization. They correspond to a sample of dijets from \textsc{Pythia} with detector-level invariant masses in the range of $23.0 < M < 28.0$~GeV/$c^2$.\par

As can be seen from the plot, the separation in $x$ between the two distributions increases with the rising sum $\eta_3 + \eta_4$, indicating a larger momentum asymmetry of the colliding partons. Similar distributions for Barrel-Barrel dijet topology provide sensitivity down to $x \sim 0.05$. It is evident that extending the measurement into the Endcap region grants access to significantly lower values of $x$. Additionally, the substantial imbalance in the initial state momentum fractions, coupled with the shapes of well-established unpolarized PDFs, suggests that the low-$x$ peak is predominantly governed by gluons, while the high-$x$ partons are primarily valence quarks~\cite{PhysRevD.86.094009}.

Figure \ref{fig:ALL_DiffTopo} presents our new $A_{LL}$ results as a function of dijet mass from the 2015 dataset, depicted in green, compared with previous results, shown in gray. These results are categorized according to the same event topologies as used in Figs.~\ref{fig:Dijet_subprocess_nloct18} and ~\ref{fig:Dijet_Kinematics}. The $A_{LL}$ data points are positioned at the mass-weighted average of each dijet mass bin and have been corrected back to the parton level. The error bars show the statistical uncertainties. The height of the uncertainty boxes represents the total systematic uncertainty, incorporating contributions from trigger and reconstruction bias, residual transverse polarization components in the beams, and underlying event uncertainties. The relative luminosity uncertainty, common to all points, is indicated by the small gray band on the horizontal axis. An overall vertical scale uncertainty (of $6.1\%$ for 2015 and $6.5\%$ for 2009), due to the uncertainty in the absolute beam polarizations, is not displayed. The width of the uncertainty boxes include the total systematic uncertainty associated with the corrected dijet invariant mass values. This includes uncertainties in dijet invariant mass correction, calorimeter tower gains and efficiencies, TPC momentum resolution, tracking efficiencies, and additional uncertainty related to differences among predictions from the \textsc{Pythia} tune sets. The total systematic uncertainty also incorporates the study of UE effects in both simulation and data. The comparison between the new results and the previous measurements reveals good agreement, with improved statistical precision.\par

The $A_{LL}$ asymmetry results presented in the Fig.~\ref{fig:ALL_DiffTopo} are also compared with two different theoretical model predictions. The next-to-leading order (NLO) dijet calculations were performed using the anti-$k_T$ algorithm with a jet radius parameter of $R = 0.6$, as implemented in the dijet production code by de Florian \textit{et al.}~\cite{deFlorian:1998qp}. This choice is consistent with our previous measurements at $\sqrt{s} = 200$ GeV. The anti-$k_T$ algorithm is less sensitive to diffuse soft backgrounds from the underlying event and pileup, thereby significantly reducing trigger bias. The radius parameter $R = 0.6$ is selected to maximize jet reconstruction efficiency while minimizing sensitivity to differences in quark- and gluon-jet fragmentation. Theoretical predictions were generated using the polarized PDF sets DSSV2014~\cite{ref:DSSV2014} and NNPDFpol1.1~\cite{ref:NNPDF}, with MRST2008~\cite{Martin:2009iq} and NNPDF2.3~\cite{Ball:2013hta} used as the corresponding unpolarized PDFs in the denominator of the asymmetry calculations. Although the jet radius is matched between the data and theory, previous studies from both experimental and theoretical sides have demonstrated that the double-spin asymmetry $A_{LL}$ is largely insensitive to the choice of $R$, as long as it is not too small. The renormalization scale was set to the jet transverse momentum, $\mu_R = p_{\text{T,jet}}$. To estimate the associated systematic uncertainty, the scale was varied between $p_{\text{T,jet}}/2$ and $2p_{\text{T,jet}}$. This scale uncertainty was found to be small compared to the PDF uncertainty for both jet and dijet $A_{LL}$ measurements. The data are generally in good agreement with current theoretical model expectations. This suggests that incorporating these results into global analyses may not significantly alter the integrated value of $\Delta g(x)$ but would lead to reduced uncertainties.\par

\subsection{Comparison with JAM22 calculations}

The global analysis from the JAM collaboration contains both positive and negative gluon polarization results. The latter does not impose positivity constraints, which are
\begin{equation}
|\Delta f_{i} (x,Q^{2})| < f_{i} (x,Q^{2})
\end{equation}
where $f_i$ represents the various quark, antiquark, and gluon flavors. The positivity bound on the helicity can be derived naturally from their definitions in terms of the probabilistic interpretation and cross-section asymmetries~\cite{ref:positivity_Altarelli:1998gn}. The violation of the positivity bounds could exhibit hard processes with unacceptable negative cross-sections, for example, the Higgs boson production~\cite{deFlorian:2024utd}. If we nonetheless allow the cross section to be negative and implement the JAM22 PDF sets into the NLO polarized cross section calculations~\cite{ref:Frixione:1995ms,ref:Frixione:1997ks,ref:Frixione:1997np,deFlorian:1998qp}, the theoretical expectations for $A_{LL}$ 
are presented in Fig.~\ref{fig:ALL_withJAM22}. Here we use the same NLO dijet code as in Fig.~\ref{fig:ALL_DiffTopo}, and all the computational settings remain consistent except for the choice of PDFs. The JAM22 PDF sets feature 101 and 91 replicas for positive and negative gluon polarization, respectively. The corresponding asymmetries, depicted in blue and red for these distinct PDF sets and their replicas, are presented. Additionally, specific results for varied sub-processes are demonstrated for central replica (ID = 0), where dashed, dotted, dash-dotted, and solid lines represent the separate contributions from gluon-gluon, quark-gluon, and quark-quark scattering processes and their combined effect, respectively. In scenarios of negative gluon polarization, the positive asymmetry from the gluon-gluon scattering process offsets the negative asymmetry from quark-gluon scattering, ensuring a final asymmetry above zero. Remarkably, for this substantial asymmetry in the gluon-gluon process, the polarized cross-section surpasses the unpolarized counterpart at high dijet mass values, leading to negative $\sigma_{+-}$ values for gluon-gluon scattering in these regions. Specifically, for these dijet topologies, $\sigma_{+-}$ is found to be negative when the dijet invariant mass exceeds 70 GeV$/c^{2}$, 50 GeV$/c^{2}$, and 40 GeV$/c^{2}$, from top to bottom panels, respectively.\par

The data from 2009 and 2015 are combined by average weighting and then compared with the theoretical expectations from JAM22 in Fig.~\ref{fig:ALL_withJAM22}. In order to provide a quantitative comparison, a $\chi^{2}$ test between the data and theoretical expectations was calculated following the discussion in~\cite{ref:Stump:2001gu}.\par

There are three correlated systematic uncertainties in our results. One is associated with the relative luminosity and is estimated to be $\beta_{1} = 0.00065$ for 2015 and 0.0005 for 2009 measurement. The second one is associated with the dijet energy scale. This uncertainty is estimated as $\beta_{2} = \frac{\partial A_{LL}}{\partial M} \times \Delta M$. The differential factor is from a third order polynomial fit of the theory curve, and $\Delta M$ is the systematic uncertainty on the dijet invariant mass. The last one is the scale uncertainty from polarization, and is estimated to be $\beta_{3} = 6.1\%$ for 2015 and $6.5\%$ for 2009. This polarization uncertainty scales both the $A_{LL}$ and its uncertainty $\sigma_{A_{LL}}$, thus the expression for $\chi^{2}$ is:
\begin{widetext}
\begin{equation}
\chi^{2} = \sum_{i=1}^{N} (\frac{(1+\beta_{3,i}r_{3})m_{i}-t_{i}-\beta_{1,i}r_{1}-\beta_{2,i}r_{2}}{(1+\beta_{3,i}r_{3})\sigma_{i}})^{2} + r_{1}^{2} + r_{2}^{2} + r_{3}^{2}
\end{equation}
\end{widetext}
where $m_{i}$ is the measured data asymmetry; $t_{i}$ is the theory expectation at the mass of the data point; $\sigma_{i}$ is the uncertainty on the measured data asymmetry, and it is the quadrature sum of the statistical and point-to-point systematic uncertainties. $r_{1}, r_{2}$ and $ r_{3}$ are the corresponding free parameters associated with the three correlated uncertainties, which are chosen to minimize the $\chi^{2}$ values.\par

$\chi^{2}$ was calculated for each of the replicas. The average $\chi^{2}$ for the 101 replicas with positive gluon polarization is 20.4 for 20 degrees of freedom, and it is 32.7 for negative gluon polarization. Our results disfavor the negative gluon polarization fit at the 3.5$\sigma$ level.\par

\section{Summary}\label{sec:Conclusion}

The STAR collaboration presented high-precision longitudinal double-spin asymmetry measurements for dijets at intermediate pseudorapidities, utilizing data from polarized $pp$ collisions at a center-of-mass energy of 200 GeV collected in 2015. These $A_{LL}$ results exhibit good agreement with previous measurements, underscoring the robustness and consistency of the experimental approach and analysis. They are also consistent with the predictions of previous global analysis fits that find the gluon polarization to be positive. However, they strongly disfavor the JAM negative gluon polarization fit. These results provide insights and impose new constraints on both the magnitude and shape of the gluon helicity distribution ($\Delta g(x)$), advancing our comprehension of gluon polarization within the proton.\par

\begin{acknowledgments}

\end{acknowledgments}
We thank D. de Florian and R. Sassot for useful discussions regarding the NLO dijet $A_{LL}$ calculations.

We thank the RHIC Operations Group and SDCC at BNL, the NERSC Center at LBNL, and the Open Science Grid consortium for providing resources and support.  This work was supported in part by the Office of Nuclear Physics within the U.S. DOE Office of Science, the U.S. National Science Foundation, National Natural Science Foundation of China, Chinese Academy of Science, the Ministry of Science and Technology of China and the Chinese Ministry of Education, NSTC Taipei, the National Research Foundation of Korea, Czech Science Foundation and Ministry of Education, Youth and Sports of the Czech Republic, Hungarian National Research, Development and Innovation Office, New National Excellency Programme of the Hungarian Ministry of Human Capacities, Department of Atomic Energy and Department of Science and Technology of the Government of India, the National Science Centre and WUT ID-UB of Poland, the Ministry of Science, Education and Sports of the Republic of Croatia, German Bundesministerium f\"ur Bildung, Wissenschaft, Forschung and Technologie (BMBF), Helmholtz Association, Ministry of Education, Culture, Sports, Science, and Technology (MEXT), Japan Society for the Promotion of Science (JSPS) and Agencia Nacional de Investigaci\'on y Desarrollo (ANID) of Chile.

\appendix*

\section{Correlation matrix}
The correlation matrices for the quadrature sum of the point-to-point statistical and systematic uncertainties between the 2015 dijet results at intermediate pseudorapidity to the previous measurements of $A_{LL}$ for inclusive jets and dijets at mid-rapidity \cite{ref:STAR_2015results_ALL} are shown in Tab.~\ref{table:BE_IncJet_Corr_Matrix}~\ref{table:BE_BB_Dijet_Corr_Matrix}~\ref{table:BE_InCorr_Matrix}. Bins 1-11 represent the 11 jet $p_{\mathrm{T}}$ points that have $|\eta_{Jet}| < 0.5$ while bins 12-22 represent the 11 $p_{\mathrm{T}}$ points that have $0.5 < |\eta_{Jet}| < 1.0$. Bins 23-29 represent the 7 dijet invariant mass points from the same-sign topology ($Sign(\eta_{1}) = Sign(\eta_{2})$) and bins 30-36 represent the 7 dijet points from the opposite-sign topology ($Sign(\eta_{1}) \neq Sign(\eta_{2})$). Bins 37-43 represent the dijet in which one jet has $\eta_{Jet} < 0$ while the other jet has $0.8 < \eta_{Jet} < 1.8$; Bins 44-50 represent the dijet with one jet $0 < \eta_{Jet} < 0.8$ and the other jet $0.8 < \eta_{Jet} < 1.8$. Bins 51-56 represent the dijet with both jets $0.8 < \eta_{Jet} < 1.8$. Table~\ref{table:BE_IncJet_Corr_Matrix} gives the correlations between the mid-rapidity inclusive jet and the intermediate pseudorapidity dijet results. Table~\ref{table:BE_BB_Dijet_Corr_Matrix} gives the correlations between the mid-rapidity dijet and the intermediate pseudorapidity dijet results. Table~\ref{table:BE_InCorr_Matrix} shows the correlations within the intermediate pseudorapidity measurements.\par

There are two systematic uncertainties on $A_{LL}$ that are not included in the tables, but are $100\%$ correlated between bins: the relative luminosity uncertainty and the polarization uncertainty. The relative luminosity uncertainty is a vertical shift uncertainty with magnitude of 0.0007 and the polarization uncertainty is a vertical scale uncertainty of $6.1\%$.\par

\setlength{\tabcolsep}{6pt}
\begin{sidewaystable}
\setlength{\tabcolsep}{6pt}
\centering
\caption{Correlation matrix, Column: bins 1-11 represent the inclusive jet with $|\eta_{Jet}| < 0.5$; bins 12-22 represent the inclusive jet with $0.5 < |\eta_{Jet}| < 1.0$. Row: bins 37-43 represent the intermediate pseudorapidity dijet with one jet in $-0.8 < \eta_{Jet} < 0$; bins 44-50 represent the intermediate pseudorapidity dijet with one jet in $0 < \eta_{Jet} < 0.8$; bins 51-56 represent the dijet with both jets in $0.8 < \eta_{Jet} < 1.8$.}
\resizebox{\textwidth}{!}{
\setlength{\tabcolsep}{6pt}
\begin{tabular}{c|c|c|c|c|c|c|c|c|c|c|c|c|c|c|c|c|c|c|c|c|c|c|c}
Bin & 1 & 2 & 3 & 4 & 5 & 6 & 7 & 8 & 9 & 10 & 11 & 12 & 13 & 14 & 15 & 16 & 17 & 18 & 19 & 20 & 21 & 22 & Bin\\ \hline
37 & 0.009 & 0.050 & 0.020 & 0.003 & 0.000 & 0.000 & 0.000 & 0.000 & 0.000 & 0.000 & 0.000 & 0.005 & 0.024 & 0.014 & 0.004 & 0.000 & 0.000 & 0.000 & 0.000 & 0.000 & 0.000 & 0.000 & 37 \\
38 & 0.012 & 0.057 & 0.057 & 0.035 & 0.010 & 0.001 & 0.000 & 0.000 & 0.000 & 0.000 & 0.000 & 0.008 & 0.040 & 0.032 & 0.020 & 0.008 & 0.002 & 0.000 & 0.000 & 0.000 & 0.000 & 0.000 & 38 \\
39 & 0.004 & 0.027 & 0.047 & 0.061 & 0.041 & 0.015 & 0.003 & 0.000 & 0.000 & 0.000 & 0.000 & 0.005 & 0.030 & 0.039 & 0.040 & 0.026 & 0.012 & 0.004 & 0.001 & 0.000 & 0.000 & 0.000 & 39 \\
40 & 0.000 & 0.005 & 0.017 & 0.046 & 0.065 & 0.045 & 0.016 & 0.003 & 0.000 & 0.000 & 0.000 & 0.001 & 0.010 & 0.023 & 0.042 & 0.045 & 0.029 & 0.013 & 0.005 & 0.001 & 0.000 & 0.000 & 40 \\
41 & 0.000 & 0.000 & 0.002 & 0.012 & 0.042 & 0.070 & 0.051 & 0.016 & 0.003 & 0.000 & 0.000 & 0.000 & 0.002 & 0.006 & 0.020 & 0.044 & 0.051 & 0.032 & 0.014 & 0.004 & 0.001 & 0.000 & 41 \\
42 & 0.000 & 0.000 & 0.000 & 0.001 & 0.006 & 0.032 & 0.075 & 0.082 & 0.042 & 0.010 & 0.002 & 0.000 & 0.000 & 0.001 & 0.003 & 0.016 & 0.046 & 0.072 & 0.063 & 0.031 & 0.011 & 0.003 & 42 \\
43 & 0.000 & 0.000 & 0.000 & 0.000 & 0.000 & 0.000 & 0.002 & 0.016 & 0.059 & 0.070 & 0.033 & 0.000 & 0.000 & 0.000 & 0.000 & 0.000 & 0.001 & 0.008 & 0.043 & 0.086 & 0.073 & 0.034 & 43 \\
44 & 0.010 & 0.046 & 0.032 & 0.012 & 0.002 & 0.000 & 0.000 & 0.000 & 0.000 & 0.000 & 0.000 & 0.014 & 0.062 & 0.059 & 0.034 & 0.009 & 0.001 & 0.000 & 0.000 & 0.000 & 0.000 & 0.000 & 44 \\
45 & 0.006 & 0.033 & 0.048 & 0.047 & 0.024 & 0.006 & 0.001 & 0.000 & 0.000 & 0.000 & 0.000 & 0.004 & 0.027 & 0.046 & 0.058 & 0.041 & 0.016 & 0.003 & 0.000 & 0.000 & 0.000 & 0.000 & 45 \\
46 & 0.001 & 0.009 & 0.026 & 0.054 & 0.057 & 0.030 & 0.008 & 0.001 & 0.000 & 0.000 & 0.000 & 0.001 & 0.006 & 0.017 & 0.042 & 0.059 & 0.043 & 0.017 & 0.005 & 0.001 & 0.000 & 0.000 & 46 \\
47 & 0.000 & 0.001 & 0.004 & 0.022 & 0.056 & 0.068 & 0.035 & 0.009 & 0.002 & 0.000 & 0.000 & 0.000 & 0.001 & 0.003 & 0.012 & 0.035 & 0.060 & 0.046 & 0.018 & 0.005 & 0.001 & 0.000 & 47 \\
48 & 0.000 & 0.000 & 0.000 & 0.002 & 0.015 & 0.054 & 0.079 & 0.042 & 0.009 & 0.002 & 0.001 & 0.000 & 0.000 & 0.001 & 0.002 & 0.008 & 0.027 & 0.058 & 0.049 & 0.018 & 0.005 & 0.001 & 48 \\
49 & 0.000 & 0.000 & 0.000 & 0.000 & 0.001 & 0.008 & 0.040 & 0.092 & 0.075 & 0.029 & 0.007 & 0.000 & 0.000 & 0.000 & 0.000 & 0.002 & 0.006 & 0.019 & 0.052 & 0.062 & 0.033 & 0.009 & 49 \\
50 & 0.000 & 0.000 & 0.000 & 0.000 & 0.000 & 0.000 & 0.000 & 0.002 & 0.019 & 0.069 & 0.054 & 0.000 & 0.000 & 0.000 & 0.000 & 0.000 & 0.000 & 0.001 & 0.003 & 0.013 & 0.033 & 0.040 & 50 \\
51 & 0.000 & 0.000 & 0.000 & 0.000 & 0.000 & 0.000 & 0.000 & 0.000 & 0.000 & 0.000 & 0.000 & 0.007 & 0.026 & 0.024 & 0.015 & 0.004 & 0.000 & 0.000 & 0.000 & 0.000 & 0.000 & 0.000 & 51 \\
52 & 0.000 & 0.000 & 0.000 & 0.000 & 0.000 & 0.000 & 0.000 & 0.000 & 0.000 & 0.000 & 0.000 & 0.002 & 0.012 & 0.021 & 0.026 & 0.020 & 0.010 & 0.002 & 0.000 & 0.000 & 0.000 & 0.000 & 52 \\
53 & 0.000 & 0.000 & 0.000 & 0.000 & 0.000 & 0.000 & 0.000 & 0.000 & 0.000 & 0.000 & 0.000 & 0.000 & 0.002 & 0.007 & 0.019 & 0.027 & 0.023 & 0.011 & 0.004 & 0.001 & 0.000 & 0.000 & 53 \\
54 & 0.000 & 0.000 & 0.000 & 0.000 & 0.000 & 0.000 & 0.000 & 0.000 & 0.000 & 0.000 & 0.000 & 0.000 & 0.000 & 0.001 & 0.004 & 0.014 & 0.027 & 0.026 & 0.013 & 0.004 & 0.001 & 0.000 & 54 \\
55 & 0.000 & 0.000 & 0.000 & 0.000 & 0.000 & 0.000 & 0.000 & 0.000 & 0.000 & 0.000 & 0.000 & 0.000 & 0.000 & 0.000 & 0.000 & 0.002 & 0.010 & 0.026 & 0.027 & 0.014 & 0.003 & 0.001 & 55 \\
56 & 0.000 & 0.000 & 0.000 & 0.000 & 0.000 & 0.000 & 0.000 & 0.000 & 0.000 & 0.000 & 0.000 & 0.000 & 0.000 & 0.000 & 0.000 & 0.000 & 0.000 & 0.000 & 0.000 & 0.000 & 0.000 & 0.000 & 56 \\
\hline
Bin & 1 & 2 & 3 & 4 & 5 & 6 & 7 & 8 & 9 & 10 & 11 & 12 & 13 & 14 & 15 & 16 & 17 & 18 & 19 & 20 & 21 & 22 & Bin\\
\end{tabular}}
\label{table:BE_IncJet_Corr_Matrix}
\end{sidewaystable}

\newpage
\setlength{\tabcolsep}{11pt}
\begin{sidewaystable}
\setlength{\tabcolsep}{11pt}
\centering
\caption{Correlation matrix, Column: bins 23-29 represent the 7 mid-rapidity dijet invariant mass points from the same-sign topology ($Sign(\eta_{1}) = Sign(\eta_{2})$) and bins 30-36 represent the 7 dijet points from the opposite-sign topology ($Sign(\eta_{1}) \neq Sign(\eta_{2})$). Row: bins 37-43 represent the intermediate pseudorapidity dijet with one jet in $\eta_{Jet} < 0$; bins 44-50 represent the intermediate pseudorapidity dijet with one jet in $0 < \eta_{Jet} < 0.8$; bins 51-55 represent the dijet with both jets in $0.8 < \eta_{Jet} < 1.8$.}
\resizebox{\textwidth}{!}{
\setlength{\tabcolsep}{11pt}
\begin{tabular}{c|c|c|c|c|c|c|c|c|c|c|c|c|c|c|c}
Bin & 23 & 24 & 25 & 26 & 27 & 28 & 29 & 30 & 31 & 32 & 33 & 34 & 35 & 36 & Bin\\ \hline
37 & 0.000 & 0.000 & 0.000 & 0.000 & 0.000 & 0.000 & 0.000 & 0.000 & 0.000 & 0.000 & 0.000 & 0.000 & 0.000 & 0.000 & 37 \\
38 & 0.000 & 0.000 & 0.000 & 0.000 & 0.000 & 0.000 & 0.000 & 0.000 & 0.000 & 0.000 & 0.000 & 0.000 & 0.000 & 0.000 & 38 \\
39 & 0.000 & 0.000 & 0.000 & 0.000 & 0.000 & 0.000 & 0.000 & 0.000 & 0.000 & 0.001 & 0.000 & 0.000 & 0.000 & 0.000 & 39 \\
40 & 0.000 & 0.000 & 0.000 & 0.000 & 0.000 & 0.000 & 0.000 & 0.000 & 0.000 & 0.000 & 0.000 & 0.000 & 0.000 & 0.000 & 40 \\
41 & 0.000 & 0.000 & 0.000 & 0.000 & 0.000 & 0.000 & 0.000 & 0.000 & 0.000 & 0.000 & 0.000 & 0.000 & 0.000 & 0.000 & 41 \\
42 & 0.000 & 0.000 & 0.000 & 0.000 & 0.000 & 0.000 & 0.000 & 0.000 & 0.000 & 0.000 & 0.000 & 0.000 & 0.000 & 0.000 & 42 \\
43 & 0.000 & 0.000 & 0.000 & 0.000 & 0.000 & 0.000 & 0.000 & 0.000 & 0.000 & 0.000 & 0.000 & 0.000 & 0.000 & 0.000 & 43 \\
44 & 0.000 & 0.000 & 0.000 & 0.000 & 0.000 & 0.000 & 0.000 & 0.000 & 0.000 & 0.000 & 0.000 & 0.000 & 0.000 & 0.000 & 44 \\
45 & 0.000 & 0.001 & 0.000 & 0.000 & 0.000 & 0.000 & 0.000 & 0.000 & 0.001 & 0.000 & 0.000 & 0.000 & 0.000 & 0.000 & 45 \\
46 & 0.000 & 0.000 & 0.000 & 0.000 & 0.000 & 0.000 & 0.000 & 0.000 & 0.000 & 0.000 & 0.000 & 0.000 & 0.000 & 0.000 & 46 \\
47 & 0.000 & 0.000 & 0.000 & 0.000 & 0.000 & 0.000 & 0.000 & 0.000 & 0.000 & 0.000 & 0.000 & 0.000 & 0.000 & 0.000 & 47 \\
48 & 0.000 & 0.000 & 0.000 & 0.000 & 0.000 & 0.000 & 0.000 & 0.000 & 0.000 & 0.000 & 0.000 & 0.000 & 0.000 & 0.000 & 48 \\
49 & 0.000 & 0.000 & 0.000 & 0.000 & 0.000 & 0.000 & 0.000 & 0.000 & 0.000 & 0.000 & 0.000 & 0.000 & 0.000 & 0.000 & 49 \\
50 & 0.000 & 0.000 & 0.000 & 0.000 & 0.000 & 0.000 & 0.000 & 0.000 & 0.000 & 0.000 & 0.000 & 0.000 & 0.000 & 0.000 & 50 \\
51 & 0.000 & 0.000 & 0.000 & 0.000 & 0.000 & 0.000 & 0.000 & 0.000 & 0.000 & 0.000 & 0.000 & 0.000 & 0.000 & 0.000 & 51 \\
52 & 0.000 & 0.000 & 0.000 & 0.000 & 0.000 & 0.000 & 0.000 & 0.000 & 0.000 & 0.000 & 0.000 & 0.000 & 0.000 & 0.000 & 52 \\
53 & 0.000 & 0.000 & 0.000 & 0.000 & 0.000 & 0.000 & 0.000 & 0.000 & 0.000 & 0.000 & 0.000 & 0.000 & 0.000 & 0.000 & 53 \\
54 & 0.000 & 0.000 & 0.000 & 0.000 & 0.000 & 0.000 & 0.000 & 0.000 & 0.000 & 0.000 & 0.000 & 0.000 & 0.000 & 0.000 & 54 \\
55 & 0.000 & 0.000 & 0.000 & 0.000 & 0.000 & 0.000 & 0.000 & 0.000 & 0.000 & 0.000 & 0.000 & 0.000 & 0.000 & 0.000 & 55 \\
56 & 0.000 & 0.000 & 0.000 & 0.000 & 0.000 & 0.005 & 0.000 & 0.000 & 0.000 & 0.000 & 0.000 & 0.000 & 0.007 & 0.000 & 56 \\
\hline
Bin & 23 & 24 & 25 & 26 & 27 & 28 & 29 & 30 & 31 & 32 & 33 & 34 & 35 & 36 & Bin\\
\end{tabular}}
\label{table:BE_BB_Dijet_Corr_Matrix}
\end{sidewaystable}

\setlength{\tabcolsep}{6pt}
\begin{sidewaystable}
\setlength{\tabcolsep}{6pt}
\centering
\caption{Correlation matrix: bins 37-43 represent the intermediate pseudorapidity dijet with one jet in $\eta_{Jet} < 0$; bins 44-50 represent the intermediate pseudorapidity dijet with one jet in $0 < \eta_{Jet} < 0.8$; bins 51-55 represent the dijet with both jets in $0.8 < \eta_{Jet} < 1.8$.}
\resizebox{\textwidth}{!}{
\setlength{\tabcolsep}{6pt}
\begin{tabular}{c|c|c|c|c|c|c|c|c|c|c|c|c|c|c|c|c|c|c|c|c|c}
Bin & 37 & 38 & 39 & 40 & 41 & 42 & 43 & 44 & 45 & 46 & 47 & 48 & 49 & 50 & 51 & 52 & 53 & 54 & 55 & 56 & Bin\\ \hline
37 & 1.000 & 0.011 & 0.014 & 0.013 & 0.015 & 0.010 & 0.007 & 0.017 & 0.013 & 0.015 & 0.021 & 0.014 & 0.013 & 0.007 & 0.012 & 0.013 & 0.016 & 0.027 & 0.012 & 0.001 & 37 \\
38 & 0.011 & 1.000 & 0.014 & 0.012 & 0.015 & 0.010 & 0.007 & 0.017 & 0.013 & 0.015 & 0.021 & 0.014 & 0.013 & 0.006 & 0.012 & 0.013 & 0.015 & 0.026 & 0.011 & 0.001 & 38 \\
39 & 0.014 & 0.014 & 1.000 & 0.015 & 0.018 & 0.012 & 0.008 & 0.020 & 0.015 & 0.018 & 0.026 & 0.016 & 0.016 & 0.008 & 0.014 & 0.015 & 0.019 & 0.032 & 0.014 & 0.001 & 39 \\
40 & 0.013 & 0.012 & 0.015 & 1.000 & 0.016 & 0.011 & 0.007 & 0.019 & 0.014 & 0.016 & 0.024 & 0.015 & 0.015 & 0.007 & 0.013 & 0.014 & 0.017 & 0.029 & 0.013 & 0.001 & 40 \\
41 & 0.015 & 0.015 & 0.018 & 0.016 & 1.000 & 0.013 & 0.009 & 0.022 & 0.017 & 0.019 & 0.028 & 0.018 & 0.017 & 0.009 & 0.016 & 0.017 & 0.020 & 0.034 & 0.015 & 0.001 & 41 \\
42 & 0.010 & 0.010 & 0.012 & 0.011 & 0.013 & 1.000 & 0.006 & 0.015 & 0.012 & 0.013 & 0.019 & 0.012 & 0.012 & 0.006 & 0.011 & 0.011 & 0.014 & 0.024 & 0.010 & 0.012 & 42 \\
43 & 0.007 & 0.007 & 0.008 & 0.007 & 0.009 & 0.006 & 1.000 & 0.010 & 0.008 & 0.009 & 0.013 & 0.008 & 0.008 & 0.004 & 0.007 & 0.007 & 0.009 & 0.016 & 0.007 & 0.001 & 43 \\
44 & 0.017 & 0.017 & 0.020 & 0.019 & 0.022 & 0.015 & 0.010 & 1.000 & 0.019 & 0.022 & 0.032 & 0.021 & 0.020 & 0.010 & 0.018 & 0.019 & 0.023 & 0.040 & 0.017 & 0.001 & 44 \\
45 & 0.013 & 0.013 & 0.015 & 0.014 & 0.017 & 0.012 & 0.008 & 0.019 & 1.000 & 0.017 & 0.024 & 0.016 & 0.015 & 0.007 & 0.014 & 0.015 & 0.018 & 0.030 & 0.013 & 0.001 & 45 \\
46 & 0.015 & 0.015 & 0.018 & 0.016 & 0.019 & 0.013 & 0.009 & 0.022 & 0.017 & 1.000 & 0.027 & 0.018 & 0.017 & 0.008 & 0.015 & 0.016 & 0.020 & 0.034 & 0.015 & 0.001 & 46 \\
47 & 0.021 & 0.021 & 0.026 & 0.024 & 0.028 & 0.019 & 0.013 & 0.032 & 0.024 & 0.027 & 1.000 & 0.026 & 0.025 & 0.012 & 0.022 & 0.024 & 0.029 & 0.050 & 0.022 & 0.002 & 47 \\
48 & 0.014 & 0.014 & 0.016 & 0.015 & 0.018 & 0.012 & 0.008 & 0.021 & 0.016 & 0.018 & 0.026 & 1.000 & 0.016 & 0.008 & 0.014 & 0.015 & 0.019 & 0.032 & 0.014 & 0.001 & 48 \\
49 & 0.013 & 0.013 & 0.016 & 0.015 & 0.017 & 0.012 & 0.008 & 0.020 & 0.015 & 0.017 & 0.025 & 0.016 & 1.000 & 0.008 & 0.014 & 0.015 & 0.018 & 0.031 & 0.013 & 0.007 & 49 \\
50 & 0.007 & 0.006 & 0.008 & 0.007 & 0.009 & 0.006 & 0.004 & 0.010 & 0.007 & 0.008 & 0.012 & 0.008 & 0.008 & 1.000 & 0.007 & 0.007 & 0.009 & 0.015 & 0.007 & 0.000 & 50 \\
51 & 0.012 & 0.012 & 0.014 & 0.013 & 0.016 & 0.011 & 0.007 & 0.018 & 0.014 & 0.015 & 0.022 & 0.014 & 0.014 & 0.007 & 1.000 & 0.013 & 0.016 & 0.028 & 0.012 & 0.001 & 51 \\
52 & 0.013 & 0.013 & 0.015 & 0.014 & 0.017 & 0.011 & 0.007 & 0.019 & 0.015 & 0.016 & 0.024 & 0.015 & 0.015 & 0.007 & 0.013 & 1.000 & 0.017 & 0.029 & 0.013 & 0.001 & 52 \\
53 & 0.016 & 0.015 & 0.019 & 0.017 & 0.020 & 0.014 & 0.009 & 0.023 & 0.018 & 0.020 & 0.029 & 0.019 & 0.018 & 0.009 & 0.016 & 0.017 & 1.000 & 0.036 & 0.016 & 0.001 & 53 \\
54 & 0.027 & 0.026 & 0.032 & 0.029 & 0.034 & 0.024 & 0.016 & 0.040 & 0.030 & 0.034 & 0.050 & 0.032 & 0.031 & 0.015 & 0.028 & 0.029 & 0.036 & 1.000 & 0.027 & 0.002 & 54 \\
55 & 0.012 & 0.011 & 0.014 & 0.013 & 0.015 & 0.010 & 0.007 & 0.017 & 0.013 & 0.015 & 0.022 & 0.014 & 0.013 & 0.007 & 0.012 & 0.013 & 0.016 & 0.027 & 1.000 & 0.001 & 55 \\
56 & 0.001 & 0.001 & 0.001 & 0.001 & 0.001 & 0.012 & 0.001 & 0.001 & 0.001 & 0.001 & 0.002 & 0.001 & 0.007 & 0.000 & 0.001 & 0.001 & 0.001 & 0.002 & 0.001 & 1.000 & 56 \\
\hline
Bin & 37 & 38 & 39 & 40 & 41 & 42 & 43 & 44 & 45 & 46 & 47 & 48 & 49 & 50 & 51 & 52 & 53 & 54 & 55 & 56 & Bin\\
\end{tabular}}
\label{table:BE_InCorr_Matrix}
\end{sidewaystable}

\bibliography{main}

\begin{thebibliography}{61}%
\makeatletter
\providecommand \@ifxundefined [1]{%
 \@ifx{#1\undefined}
}%
\providecommand \@ifnum [1]{%
 \ifnum #1\expandafter \@firstoftwo
 \else \expandafter \@secondoftwo
 \fi
}%
\providecommand \@ifx [1]{%
 \ifx #1\expandafter \@firstoftwo
 \else \expandafter \@secondoftwo
 \fi
}%
\providecommand \natexlab [1]{#1}%
\providecommand \enquote  [1]{``#1''}%
\providecommand \bibnamefont  [1]{#1}%
\providecommand \bibfnamefont [1]{#1}%
\providecommand \citenamefont [1]{#1}%
\providecommand \href@noop [0]{\@secondoftwo}%
\providecommand \href [0]{\begingroup \@sanitize@url \@href}%
\providecommand \@href[1]{\@@startlink{#1}\@@href}%
\providecommand \@@href[1]{\endgroup#1\@@endlink}%
\providecommand \@sanitize@url [0]{\catcode `\\12\catcode `\$12\catcode `\&12\catcode `\#12\catcode `\^12\catcode `\_12\catcode `\%12\relax}%
\providecommand \@@startlink[1]{}%
\providecommand \@@endlink[0]{}%
\providecommand \url  [0]{\begingroup\@sanitize@url \@url }%
\providecommand \@url [1]{\endgroup\@href {#1}{\urlprefix }}%
\providecommand \urlprefix  [0]{URL }%
\providecommand \Eprint [0]{\href }%
\providecommand \doibase [0]{http://dx.doi.org/}%
\providecommand \selectlanguage [0]{\@gobble}%
\providecommand \bibinfo  [0]{\@secondoftwo}%
\providecommand \bibfield  [0]{\@secondoftwo}%
\providecommand \translation [1]{[#1]}%
\providecommand \BibitemOpen [0]{}%
\providecommand \bibitemStop [0]{}%
\providecommand \bibitemNoStop [0]{.\EOS\space}%
\providecommand \EOS [0]{\spacefactor3000\relax}%
\providecommand \BibitemShut  [1]{\csname bibitem#1\endcsname}%
\let\auto@bib@innerbib\@empty
\bibitem [{\citenamefont {Jaffe}\ and\ \citenamefont {Manohar}(1990)}]{Jaffe:1989jz}%
  \BibitemOpen
  \bibfield  {author} {\bibinfo {author} {\bibfnamefont {R.~L.}\ \bibnamefont {Jaffe}}\ and\ \bibinfo {author} {\bibfnamefont {A.}~\bibnamefont {Manohar}},\ }\href {\doibase 10.1016/0550-3213(90)90506-9} {\bibfield  {journal} {\bibinfo  {journal} {Nucl. Phys. B}\ }\textbf {\bibinfo {volume} {337}},\ \bibinfo {pages} {509} (\bibinfo {year} {1990})}\BibitemShut {NoStop}%
\bibitem [{\citenamefont {Nocera}\ \emph {et~al.}(2014)\citenamefont {Nocera}, \citenamefont {Ball}, \citenamefont {Forte}, \citenamefont {Ridolfi},\ and\ \citenamefont {Rojo}}]{ref:NNPDF}%
  \BibitemOpen
  \bibfield  {author} {\bibinfo {author} {\bibfnamefont {E.~R.}\ \bibnamefont {Nocera}}, \bibinfo {author} {\bibfnamefont {R.~D.}\ \bibnamefont {Ball}}, \bibinfo {author} {\bibfnamefont {S.}~\bibnamefont {Forte}}, \bibinfo {author} {\bibfnamefont {G.}~\bibnamefont {Ridolfi}}, \ and\ \bibinfo {author} {\bibfnamefont {J.}~\bibnamefont {Rojo}} (\bibinfo {collaboration} {NNPDF}),\ }\href {\doibase 10.1016/j.nuclphysb.2014.08.008} {\bibfield  {journal} {\bibinfo  {journal} {Nucl. Phys. B}\ }\textbf {\bibinfo {volume} {887}},\ \bibinfo {pages} {276} (\bibinfo {year} {2014})},\ \Eprint {http://arxiv.org/abs/1406.5539} {arXiv:1406.5539 [hep-ph]} \BibitemShut {NoStop}%
\bibitem [{\citenamefont {de~Florian}\ \emph {et~al.}(2014)\citenamefont {de~Florian}, \citenamefont {Sassot}, \citenamefont {Stratmann},\ and\ \citenamefont {Vogelsang}}]{ref:DSSV2014}%
  \BibitemOpen
  \bibfield  {author} {\bibinfo {author} {\bibfnamefont {D.}~\bibnamefont {de~Florian}}, \bibinfo {author} {\bibfnamefont {R.}~\bibnamefont {Sassot}}, \bibinfo {author} {\bibfnamefont {M.}~\bibnamefont {Stratmann}}, \ and\ \bibinfo {author} {\bibfnamefont {W.}~\bibnamefont {Vogelsang}},\ }\href {\doibase 10.1103/PhysRevLett.113.012001} {\bibfield  {journal} {\bibinfo  {journal} {Phys. Rev. Lett.}\ }\textbf {\bibinfo {volume} {113}},\ \bibinfo {pages} {012001} (\bibinfo {year} {2014})},\ \Eprint {http://arxiv.org/abs/1404.4293} {arXiv:1404.4293 [hep-ph]} \BibitemShut {NoStop}%
\bibitem [{\citenamefont {Alekseev}\ \emph {et~al.}(2003)\citenamefont {Alekseev} \emph {et~al.}}]{Alekseev:2003sk}%
  \BibitemOpen
  \bibfield  {author} {\bibinfo {author} {\bibfnamefont {I.}~\bibnamefont {Alekseev}} \emph {et~al.},\ }\href {\doibase 10.1016/S0168-9002(02)01946-0} {\bibfield  {journal} {\bibinfo  {journal} {Nucl. Instrum. Meth. A}\ }\textbf {\bibinfo {volume} {499}},\ \bibinfo {pages} {392} (\bibinfo {year} {2003})}\BibitemShut {NoStop}%
\bibitem [{\citenamefont {Abelev}\ \emph {et~al.}(2006)\citenamefont {Abelev} \emph {et~al.}}]{ref:STAR_2006_ALL}%
  \BibitemOpen
  \bibfield  {author} {\bibinfo {author} {\bibfnamefont {B.~I.}\ \bibnamefont {Abelev}} \emph {et~al.} (\bibinfo {collaboration} {STAR}),\ }\href {\doibase 10.1103/PhysRevLett.97.252001} {\bibfield  {journal} {\bibinfo  {journal} {Phys. Rev. Lett.}\ }\textbf {\bibinfo {volume} {97}},\ \bibinfo {pages} {252001} (\bibinfo {year} {2006})},\ \Eprint {http://arxiv.org/abs/hep-ex/0608030} {arXiv:hep-ex/0608030} \BibitemShut {NoStop}%
\bibitem [{\citenamefont {Abelev}\ \emph {et~al.}(2008)\citenamefont {Abelev} \emph {et~al.}}]{ref:STAR_2007_ALL}%
  \BibitemOpen
  \bibfield  {author} {\bibinfo {author} {\bibfnamefont {B.~I.}\ \bibnamefont {Abelev}} \emph {et~al.} (\bibinfo {collaboration} {STAR}),\ }\href {\doibase 10.1103/PhysRevLett.100.232003} {\bibfield  {journal} {\bibinfo  {journal} {Phys. Rev. Lett.}\ }\textbf {\bibinfo {volume} {100}},\ \bibinfo {pages} {232003} (\bibinfo {year} {2008})},\ \Eprint {http://arxiv.org/abs/0710.2048} {arXiv:0710.2048 [hep-ex]} \BibitemShut {NoStop}%
\bibitem [{\citenamefont {Adamczyk}\ \emph {et~al.}(2012)\citenamefont {Adamczyk} \emph {et~al.}}]{ref:STAR_2012pub_ALL}%
  \BibitemOpen
  \bibfield  {author} {\bibinfo {author} {\bibfnamefont {L.}~\bibnamefont {Adamczyk}} \emph {et~al.} (\bibinfo {collaboration} {STAR}),\ }\href {\doibase 10.1103/PhysRevD.86.032006} {\bibfield  {journal} {\bibinfo  {journal} {Phys. Rev. D}\ }\textbf {\bibinfo {volume} {86}},\ \bibinfo {pages} {032006} (\bibinfo {year} {2012})},\ \Eprint {http://arxiv.org/abs/1205.2735} {arXiv:1205.2735 [nucl-ex]} \BibitemShut {NoStop}%
\bibitem [{\citenamefont {Adamczyk}\ \emph {et~al.}(2015)\citenamefont {Adamczyk} \emph {et~al.}}]{ref:Adamczyk_2015_ALL}%
  \BibitemOpen
  \bibfield  {author} {\bibinfo {author} {\bibfnamefont {L.}~\bibnamefont {Adamczyk}} \emph {et~al.} (\bibinfo {collaboration} {STAR}),\ }\href {\doibase 10.1103/physrevlett.115.092002} {\bibfield  {journal} {\bibinfo  {journal} {Phys. Rev. Lett.}\ }\textbf {\bibinfo {volume} {115}},\ \bibinfo {pages} {092002} (\bibinfo {year} {2015})},\ \Eprint {http://arxiv.org/abs/1405.5134} {arXiv:1405.5134 [hep-ex]} \BibitemShut {NoStop}%
\bibitem [{\citenamefont {Abelev}\ \emph {et~al.}(2009)\citenamefont {Abelev} \emph {et~al.}}]{ref:STAR_2009_pi0_ALL}%
  \BibitemOpen
  \bibfield  {author} {\bibinfo {author} {\bibfnamefont {B.~I.}\ \bibnamefont {Abelev}} \emph {et~al.} (\bibinfo {collaboration} {STAR}),\ }\href {\doibase 10.1103/PhysRevD.80.111108} {\bibfield  {journal} {\bibinfo  {journal} {Phys. Rev. D}\ }\textbf {\bibinfo {volume} {80}},\ \bibinfo {pages} {111108} (\bibinfo {year} {2009})},\ \Eprint {http://arxiv.org/abs/0911.2773} {arXiv:0911.2773 [hep-ex]} \BibitemShut {NoStop}%
\bibitem [{\citenamefont {Adamczyk}\ \emph {et~al.}(2014)\citenamefont {Adamczyk} \emph {et~al.}}]{ref:STAR_2006_Endcap_pi0_ALL}%
  \BibitemOpen
  \bibfield  {author} {\bibinfo {author} {\bibfnamefont {L.}~\bibnamefont {Adamczyk}} \emph {et~al.} (\bibinfo {collaboration} {STAR}),\ }\href {\doibase 10.1103/PhysRevD.89.012001} {\bibfield  {journal} {\bibinfo  {journal} {Phys. Rev. D}\ }\textbf {\bibinfo {volume} {89}},\ \bibinfo {pages} {012001} (\bibinfo {year} {2014})},\ \Eprint {http://arxiv.org/abs/1309.1800} {arXiv:1309.1800 [nucl-ex]} \BibitemShut {NoStop}%
\bibitem [{\citenamefont {Adare}\ \emph {et~al.}(2009{\natexlab{a}})\citenamefont {Adare} \emph {et~al.}}]{PHENIX:2008sgl}%
  \BibitemOpen
  \bibfield  {author} {\bibinfo {author} {\bibfnamefont {A.}~\bibnamefont {Adare}} \emph {et~al.} (\bibinfo {collaboration} {PHENIX}),\ }\href {\doibase 10.1103/PhysRevD.79.012003} {\bibfield  {journal} {\bibinfo  {journal} {Phys. Rev. D}\ }\textbf {\bibinfo {volume} {79}},\ \bibinfo {pages} {012003} (\bibinfo {year} {2009}{\natexlab{a}})},\ \Eprint {http://arxiv.org/abs/0810.0701} {arXiv:0810.0701 [hep-ex]} \BibitemShut {NoStop}%
\bibitem [{\citenamefont {Adare}\ \emph {et~al.}(2009{\natexlab{b}})\citenamefont {Adare} \emph {et~al.}}]{PHENIX:2008swq}%
  \BibitemOpen
  \bibfield  {author} {\bibinfo {author} {\bibfnamefont {A.}~\bibnamefont {Adare}} \emph {et~al.} (\bibinfo {collaboration} {PHENIX}),\ }\href {\doibase 10.1103/PhysRevLett.103.012003} {\bibfield  {journal} {\bibinfo  {journal} {Phys. Rev. Lett.}\ }\textbf {\bibinfo {volume} {103}},\ \bibinfo {pages} {012003} (\bibinfo {year} {2009}{\natexlab{b}})},\ \Eprint {http://arxiv.org/abs/0810.0694} {arXiv:0810.0694 [hep-ex]} \BibitemShut {NoStop}%
\bibitem [{\citenamefont {Adare}\ \emph {et~al.}(2014)\citenamefont {Adare} \emph {et~al.}}]{PHENIX:2014gbf}%
  \BibitemOpen
  \bibfield  {author} {\bibinfo {author} {\bibfnamefont {A.}~\bibnamefont {Adare}} \emph {et~al.} (\bibinfo {collaboration} {PHENIX}),\ }\href {\doibase 10.1103/PhysRevD.90.012007} {\bibfield  {journal} {\bibinfo  {journal} {Phys. Rev. D}\ }\textbf {\bibinfo {volume} {90}},\ \bibinfo {pages} {012007} (\bibinfo {year} {2014})},\ \Eprint {http://arxiv.org/abs/1402.6296} {arXiv:1402.6296 [hep-ex]} \BibitemShut {NoStop}%
\bibitem [{\citenamefont {de~Florian}(2009)}]{deFlorian:2009fw}%
  \BibitemOpen
  \bibfield  {author} {\bibinfo {author} {\bibfnamefont {D.}~\bibnamefont {de~Florian}},\ }\href {\doibase 10.1103/PhysRevD.79.114014} {\bibfield  {journal} {\bibinfo  {journal} {Phys. Rev. D}\ }\textbf {\bibinfo {volume} {79}},\ \bibinfo {pages} {114014} (\bibinfo {year} {2009})},\ \Eprint {http://arxiv.org/abs/0904.4402} {arXiv:0904.4402 [hep-ph]} \BibitemShut {NoStop}%
\bibitem [{\citenamefont {Adamczyk}\ \emph {et~al.}(2017)\citenamefont {Adamczyk} \emph {et~al.}}]{ref:STAR_2009_Barrel_dijet_ALL}%
  \BibitemOpen
  \bibfield  {author} {\bibinfo {author} {\bibfnamefont {L.}~\bibnamefont {Adamczyk}} \emph {et~al.} (\bibinfo {collaboration} {STAR}),\ }\href {\doibase 10.1103/PhysRevD.95.071103} {\bibfield  {journal} {\bibinfo  {journal} {Phys. Rev. D}\ }\textbf {\bibinfo {volume} {95}},\ \bibinfo {pages} {071103} (\bibinfo {year} {2017})},\ \Eprint {http://arxiv.org/abs/1610.06616} {arXiv:1610.06616 [hep-ex]} \BibitemShut {NoStop}%
\bibitem [{\citenamefont {Adam}\ \emph {et~al.}(2018{\natexlab{a}})\citenamefont {Adam} \emph {et~al.}}]{ref:STAR_2009_Endcap_ALL}%
  \BibitemOpen
  \bibfield  {author} {\bibinfo {author} {\bibfnamefont {J.}~\bibnamefont {Adam}} \emph {et~al.} (\bibinfo {collaboration} {STAR}),\ }\href {\doibase 10.1103/PhysRevD.98.032011} {\bibfield  {journal} {\bibinfo  {journal} {Phys. Rev. D}\ }\textbf {\bibinfo {volume} {98}},\ \bibinfo {pages} {032011} (\bibinfo {year} {2018}{\natexlab{a}})},\ \Eprint {http://arxiv.org/abs/1805.09742} {arXiv:1805.09742 [hep-ex]} \BibitemShut {NoStop}%
\bibitem [{\citenamefont {de~Florian}\ \emph {et~al.}(2019)\citenamefont {de~Florian}, \citenamefont {Lucero}, \citenamefont {Sassot}, \citenamefont {Stratmann},\ and\ \citenamefont {Vogelsang}}]{DeFlorian:2019xxt}%
  \BibitemOpen
  \bibfield  {author} {\bibinfo {author} {\bibfnamefont {D.}~\bibnamefont {de~Florian}}, \bibinfo {author} {\bibfnamefont {G.~A.}\ \bibnamefont {Lucero}}, \bibinfo {author} {\bibfnamefont {R.}~\bibnamefont {Sassot}}, \bibinfo {author} {\bibfnamefont {M.}~\bibnamefont {Stratmann}}, \ and\ \bibinfo {author} {\bibfnamefont {W.}~\bibnamefont {Vogelsang}},\ }\href {\doibase 10.1103/PhysRevD.100.114027} {\bibfield  {journal} {\bibinfo  {journal} {Phys. Rev. D}\ }\textbf {\bibinfo {volume} {100}},\ \bibinfo {pages} {114027} (\bibinfo {year} {2019})},\ \Eprint {http://arxiv.org/abs/1902.10548} {arXiv:1902.10548 [hep-ph]} \BibitemShut {NoStop}%
\bibitem [{\citenamefont {Adam}\ \emph {et~al.}(2019)\citenamefont {Adam} \emph {et~al.}}]{ref:STAR_2012_ALL}%
  \BibitemOpen
  \bibfield  {author} {\bibinfo {author} {\bibfnamefont {J.}~\bibnamefont {Adam}} \emph {et~al.} (\bibinfo {collaboration} {STAR}),\ }\href {\doibase 10.1103/PhysRevD.100.052005} {\bibfield  {journal} {\bibinfo  {journal} {Phys. Rev. D}\ }\textbf {\bibinfo {volume} {100}},\ \bibinfo {pages} {052005} (\bibinfo {year} {2019})},\ \Eprint {http://arxiv.org/abs/1906.02740} {arXiv:1906.02740 [hep-ex]} \BibitemShut {NoStop}%
\bibitem [{\citenamefont {Abdallah}\ \emph {et~al.}(2022{\natexlab{a}})\citenamefont {Abdallah} \emph {et~al.}}]{ref:STAR_2013_ALL}%
  \BibitemOpen
  \bibfield  {author} {\bibinfo {author} {\bibfnamefont {M.~S.}\ \bibnamefont {Abdallah}} \emph {et~al.} (\bibinfo {collaboration} {STAR}),\ }\href {\doibase 10.1103/PhysRevD.105.092011} {\bibfield  {journal} {\bibinfo  {journal} {Phys. Rev. D}\ }\textbf {\bibinfo {volume} {105}},\ \bibinfo {pages} {092011} (\bibinfo {year} {2022}{\natexlab{a}})},\ \Eprint {http://arxiv.org/abs/2110.11020} {arXiv:2110.11020 [hep-ex]} \BibitemShut {NoStop}%
\bibitem [{\citenamefont {Adam}\ \emph {et~al.}(2018{\natexlab{b}})\citenamefont {Adam} \emph {et~al.}}]{ref:STAR_FMS_pi0_ALL}%
  \BibitemOpen
  \bibfield  {author} {\bibinfo {author} {\bibfnamefont {J.}~\bibnamefont {Adam}} \emph {et~al.} (\bibinfo {collaboration} {STAR}),\ }\href {\doibase 10.1103/PhysRevD.98.032013} {\bibfield  {journal} {\bibinfo  {journal} {Phys. Rev. D}\ }\textbf {\bibinfo {volume} {98}},\ \bibinfo {pages} {032013} (\bibinfo {year} {2018}{\natexlab{b}})},\ \Eprint {http://arxiv.org/abs/1805.09745} {arXiv:1805.09745 [hep-ex]} \BibitemShut {NoStop}%
\bibitem [{\citenamefont {Adare}\ \emph {et~al.}(2016)\citenamefont {Adare} \emph {et~al.}}]{PHENIX:2015fxo}%
  \BibitemOpen
  \bibfield  {author} {\bibinfo {author} {\bibfnamefont {A.}~\bibnamefont {Adare}} \emph {et~al.} (\bibinfo {collaboration} {PHENIX}),\ }\href {\doibase 10.1103/PhysRevD.93.011501} {\bibfield  {journal} {\bibinfo  {journal} {Phys. Rev. D}\ }\textbf {\bibinfo {volume} {93}},\ \bibinfo {pages} {011501} (\bibinfo {year} {2016})},\ \Eprint {http://arxiv.org/abs/1510.02317} {arXiv:1510.02317 [hep-ex]} \BibitemShut {NoStop}%
\bibitem [{\citenamefont {Acharya}\ \emph {et~al.}(2020)\citenamefont {Acharya} \emph {et~al.}}]{ref:PHENIX_chargedPion_ALL}%
  \BibitemOpen
  \bibfield  {author} {\bibinfo {author} {\bibfnamefont {U.~A.}\ \bibnamefont {Acharya}} \emph {et~al.} (\bibinfo {collaboration} {PHENIX}),\ }\href {\doibase 10.1103/PhysRevD.102.032001} {\bibfield  {journal} {\bibinfo  {journal} {Phys. Rev. D}\ }\textbf {\bibinfo {volume} {102}},\ \bibinfo {pages} {032001} (\bibinfo {year} {2020})},\ \Eprint {http://arxiv.org/abs/2004.02681} {arXiv:2004.02681 [hep-ex]} \BibitemShut {NoStop}%
\bibitem [{\citenamefont {Acharya}\ \emph {et~al.}(2023)\citenamefont {Acharya} \emph {et~al.}}]{ref:PHENIX_direct_photon_ALL}%
  \BibitemOpen
  \bibfield  {author} {\bibinfo {author} {\bibfnamefont {U.}~\bibnamefont {Acharya}} \emph {et~al.} (\bibinfo {collaboration} {PHENIX}),\ }\href {\doibase 10.1103/PhysRevLett.130.251901} {\bibfield  {journal} {\bibinfo  {journal} {Phys. Rev. Lett.}\ }\textbf {\bibinfo {volume} {130}},\ \bibinfo {pages} {251901} (\bibinfo {year} {2023})},\ \Eprint {http://arxiv.org/abs/2202.08158} {arXiv:2202.08158 [hep-ex]} \BibitemShut {NoStop}%
\bibitem [{\citenamefont {Abdallah}\ \emph {et~al.}(2021{\natexlab{a}})\citenamefont {Abdallah} \emph {et~al.}}]{ref:STAR_2015results_ALL}%
  \BibitemOpen
  \bibfield  {author} {\bibinfo {author} {\bibfnamefont {M.~S.}\ \bibnamefont {Abdallah}} \emph {et~al.} (\bibinfo {collaboration} {STAR}),\ }\href@noop {} {\bibfield  {journal} {\bibinfo  {journal} {Phys. Rev. D}\ }\textbf {\bibinfo {volume} {103}},\ \bibinfo {pages} {L091103} (\bibinfo {year} {2021}{\natexlab{a}})},\ \Eprint {http://arxiv.org/abs/2103.05571} {arXiv:2103.05571 [hep-ex]} \BibitemShut {NoStop}%
\bibitem [{\citenamefont {Zhou}\ \emph {et~al.}(2022)\citenamefont {Zhou}, \citenamefont {Sato},\ and\ \citenamefont {Melnitchouk}}]{ref:jam_negative_gluon}%
  \BibitemOpen
  \bibfield  {author} {\bibinfo {author} {\bibfnamefont {Y.}~\bibnamefont {Zhou}}, \bibinfo {author} {\bibfnamefont {N.}~\bibnamefont {Sato}}, \ and\ \bibinfo {author} {\bibfnamefont {W.}~\bibnamefont {Melnitchouk}} (\bibinfo {collaboration} {JAM}),\ }\href {\doibase 10.1103/PhysRevD.105.074022} {\bibfield  {journal} {\bibinfo  {journal} {Phys. Rev. D}\ }\textbf {\bibinfo {volume} {105}},\ \bibinfo {pages} {074022} (\bibinfo {year} {2022})},\ \Eprint {http://arxiv.org/abs/2201.02075} {arXiv:2201.02075 [hep-ph]} \BibitemShut {NoStop}%
\bibitem [{\citenamefont {Vogelsang}(2023)}]{ref:dssv_spin2023}%
  \BibitemOpen
  \bibfield  {author} {\bibinfo {author} {\bibfnamefont {W.}~\bibnamefont {Vogelsang}},\ }in\ \href@noop {} {\emph {\bibinfo {booktitle} {{25th International Spin Physics Symposium (SPIN 2023)}}}}\ (\bibinfo {year} {2023})\BibitemShut {NoStop}%
\bibitem [{\citenamefont {Hunt-Smith}\ \emph {et~al.}(2024)\citenamefont {Hunt-Smith}, \citenamefont {Cocuzza}, \citenamefont {Melnitchouk}, \citenamefont {Sato}, \citenamefont {Thomas},\ and\ \citenamefont {White}}]{ref:Hunt-Smith:2024khs}%
  \BibitemOpen
  \bibfield  {author} {\bibinfo {author} {\bibfnamefont {N.~T.}\ \bibnamefont {Hunt-Smith}}, \bibinfo {author} {\bibfnamefont {C.}~\bibnamefont {Cocuzza}}, \bibinfo {author} {\bibfnamefont {W.}~\bibnamefont {Melnitchouk}}, \bibinfo {author} {\bibfnamefont {N.}~\bibnamefont {Sato}}, \bibinfo {author} {\bibfnamefont {A.~W.}\ \bibnamefont {Thomas}}, \ and\ \bibinfo {author} {\bibfnamefont {M.~J.}\ \bibnamefont {White}} (\bibinfo {collaboration} {JAM}),\ }\href@noop {} {\  (\bibinfo {year} {2024})},\ \Eprint {http://arxiv.org/abs/2403.08117} {arXiv:2403.08117 [hep-ph]} \BibitemShut {NoStop}%
\bibitem [{\citenamefont {Huang}\ \emph {et~al.}(2003)\citenamefont {Huang} \emph {et~al.}}]{ref:pCPolarimeter}%
  \BibitemOpen
  \bibfield  {author} {\bibinfo {author} {\bibfnamefont {H.}~\bibnamefont {Huang}} \emph {et~al.},\ }\href {\doibase 10.1016/S0375-9474(03)01068-6} {\bibfield  {journal} {\bibinfo  {journal} {Nucl. Phys. A}\ }\textbf {\bibinfo {volume} {721}},\ \bibinfo {pages} {356} (\bibinfo {year} {2003})}\BibitemShut {NoStop}%
\bibitem [{\citenamefont {Zelenski}\ \emph {et~al.}(2005)\citenamefont {Zelenski} \emph {et~al.}}]{ref:RHIC_HJet}%
  \BibitemOpen
  \bibfield  {author} {\bibinfo {author} {\bibfnamefont {A.}~\bibnamefont {Zelenski}} \emph {et~al.},\ }\href {\doibase 10.1016/j.nima.2004.08.080} {\bibfield  {journal} {\bibinfo  {journal} {Nucl. Instrum. Meth. A}\ }\textbf {\bibinfo {volume} {536}},\ \bibinfo {pages} {248} (\bibinfo {year} {2005})}\BibitemShut {NoStop}%
\bibitem [{\citenamefont {Ackermann}\ \emph {et~al.}(2003)\citenamefont {Ackermann} \emph {et~al.}}]{STAR:2002eio}%
  \BibitemOpen
  \bibfield  {author} {\bibinfo {author} {\bibfnamefont {K.~H.}\ \bibnamefont {Ackermann}} \emph {et~al.} (\bibinfo {collaboration} {STAR}),\ }\href {\doibase 10.1016/S0168-9002(02)01960-5} {\bibfield  {journal} {\bibinfo  {journal} {Nucl. Instrum. Meth. A}\ }\textbf {\bibinfo {volume} {499}},\ \bibinfo {pages} {624} (\bibinfo {year} {2003})}\BibitemShut {NoStop}%
\bibitem [{\citenamefont {Anderson}\ \emph {et~al.}(2003)\citenamefont {Anderson} \emph {et~al.}}]{ref:tpcNIM}%
  \BibitemOpen
  \bibfield  {author} {\bibinfo {author} {\bibfnamefont {M.}~\bibnamefont {Anderson}} \emph {et~al.},\ }\href {\doibase 10.1016/S0168-9002(02)01964-2} {\bibfield  {journal} {\bibinfo  {journal} {Nucl. Instrum. Meth. Phys. Res. A}\ }\textbf {\bibinfo {volume} {499}},\ \bibinfo {pages} {659} (\bibinfo {year} {2003})},\ \Eprint {http://arxiv.org/abs/nucl-ex/0301015} {arXiv:nucl-ex/0301015 [nucl-ex]} \BibitemShut {NoStop}%
\bibitem [{\citenamefont {Beddo}\ \emph {et~al.}(2003)\citenamefont {Beddo} \emph {et~al.}}]{Beddo:2002zx}%
  \BibitemOpen
  \bibfield  {author} {\bibinfo {author} {\bibfnamefont {M.}~\bibnamefont {Beddo}} \emph {et~al.} (\bibinfo {collaboration} {STAR}),\ }\href {\doibase 10.1016/S0168-9002(02)01970-8} {\bibfield  {journal} {\bibinfo  {journal} {Nucl. Instrum. Meth. A}\ }\textbf {\bibinfo {volume} {499}},\ \bibinfo {pages} {725} (\bibinfo {year} {2003})}\BibitemShut {NoStop}%
\bibitem [{\citenamefont {Allgower}\ \emph {et~al.}(2003)\citenamefont {Allgower} \emph {et~al.}}]{Allgower:2002zy}%
  \BibitemOpen
  \bibfield  {author} {\bibinfo {author} {\bibfnamefont {C.~E.}\ \bibnamefont {Allgower}} \emph {et~al.} (\bibinfo {collaboration} {STAR}),\ }\href {\doibase 10.1016/S0168-9002(02)01971-X} {\bibfield  {journal} {\bibinfo  {journal} {Nucl. Instrum. Meth. A}\ }\textbf {\bibinfo {volume} {499}},\ \bibinfo {pages} {740} (\bibinfo {year} {2003})}\BibitemShut {NoStop}%
\bibitem [{\citenamefont {Llope}\ \emph {et~al.}(2014)\citenamefont {Llope} \emph {et~al.}}]{ref:vpdNIM}%
  \BibitemOpen
  \bibfield  {author} {\bibinfo {author} {\bibfnamefont {W.~J.}\ \bibnamefont {Llope}} \emph {et~al.},\ }\href {\doibase 10.1016/j.nima.2014.04.080} {\bibfield  {journal} {\bibinfo  {journal} {Nucl. Instrum. Meth. Phys. Res. A}\ }\textbf {\bibinfo {volume} {759}},\ \bibinfo {pages} {23} (\bibinfo {year} {2014})},\ \Eprint {http://arxiv.org/abs/1403.6855} {arXiv:1403.6855 [physics.ins-det]} \BibitemShut {NoStop}%
\bibitem [{\citenamefont {Adler}\ \emph {et~al.}(2001)\citenamefont {Adler}, \citenamefont {Denisov}, \citenamefont {Garcia}, \citenamefont {Murray}, \citenamefont {Strobele},\ and\ \citenamefont {White}}]{ref:RHIC_ZDC}%
  \BibitemOpen
  \bibfield  {author} {\bibinfo {author} {\bibfnamefont {C.}~\bibnamefont {Adler}}, \bibinfo {author} {\bibfnamefont {A.}~\bibnamefont {Denisov}}, \bibinfo {author} {\bibfnamefont {E.}~\bibnamefont {Garcia}}, \bibinfo {author} {\bibfnamefont {M.~J.}\ \bibnamefont {Murray}}, \bibinfo {author} {\bibfnamefont {H.}~\bibnamefont {Strobele}}, \ and\ \bibinfo {author} {\bibfnamefont {S.~N.}\ \bibnamefont {White}},\ }\href {\doibase 10.1016/S0168-9002(01)00627-1} {\bibfield  {journal} {\bibinfo  {journal} {Nucl. Instrum. Meth. Phys. Res. A}\ }\textbf {\bibinfo {volume} {470}},\ \bibinfo {pages} {488} (\bibinfo {year} {2001})},\ \Eprint {http://arxiv.org/abs/nucl-ex/0008005} {arXiv:nucl-ex/0008005 [nucl-ex]} \BibitemShut {NoStop}%
\bibitem [{\citenamefont {Schmidke}\ \emph {et~al.}(2018)\citenamefont {Schmidke} \emph {et~al.}}]{ref:RHICPolG}%
  \BibitemOpen
  \bibfield  {author} {\bibinfo {author} {\bibfnamefont {W.~B.}\ \bibnamefont {Schmidke}} \emph {et~al.} (\bibinfo {collaboration} {RHIC Polarimetry Group}),\ }\href {https://wiki.bnl.gov/rhicspin/Results} {\bibfield  {journal} {\bibinfo  {journal} {Report No. BNL-209057-2018-TECH}\ } (\bibinfo {year} {2018})}\BibitemShut {NoStop}%
\bibitem [{\citenamefont {Sjostrand}\ \emph {et~al.}(2006)\citenamefont {Sjostrand}, \citenamefont {Mrenna},\ and\ \citenamefont {Skands}}]{ref:Pythia6}%
  \BibitemOpen
  \bibfield  {author} {\bibinfo {author} {\bibfnamefont {T.}~\bibnamefont {Sjostrand}}, \bibinfo {author} {\bibfnamefont {S.}~\bibnamefont {Mrenna}}, \ and\ \bibinfo {author} {\bibfnamefont {P.~Z.}\ \bibnamefont {Skands}},\ }\href {\doibase 10.1088/1126-6708/2006/05/026} {\bibfield  {journal} {\bibinfo  {journal} {J. High Energy Phys.}\ }\textbf {\bibinfo {volume} {05}},\ \bibinfo {pages} {026} (\bibinfo {year} {2006})},\ \Eprint {http://arxiv.org/abs/hep-ph/0603175} {arXiv:hep-ph/0603175 [hep-ph]} \BibitemShut {NoStop}%
\bibitem [{\citenamefont {Skands}(2010)}]{ref:PerugiaTunes}%
  \BibitemOpen
  \bibfield  {author} {\bibinfo {author} {\bibfnamefont {P.~Z.}\ \bibnamefont {Skands}},\ }\href {\doibase 10.1103/PhysRevD.82.074018} {\bibfield  {journal} {\bibinfo  {journal} {Phys. Rev. D}\ }\textbf {\bibinfo {volume} {82}},\ \bibinfo {pages} {074018} (\bibinfo {year} {2010})},\ \Eprint {http://arxiv.org/abs/1005.3457v5} {arXiv:1005.3457v5 [hep-ph]} \BibitemShut {NoStop}%
\bibitem [{\citenamefont {Adams}\ \emph {et~al.}(2005)\citenamefont {Adams} \emph {et~al.}}]{ref:starPion2005}%
  \BibitemOpen
  \bibfield  {author} {\bibinfo {author} {\bibfnamefont {J.}~\bibnamefont {Adams}} \emph {et~al.} (\bibinfo {collaboration} {STAR}),\ }\href {\doibase 10.1016/j.physletb.2005.04.041} {\bibfield  {journal} {\bibinfo  {journal} {Phys. Lett. B}\ }\textbf {\bibinfo {volume} {616}},\ \bibinfo {pages} {8} (\bibinfo {year} {2005})},\ \Eprint {http://arxiv.org/abs/nucl-ex/0309012} {arXiv:nucl-ex/0309012 [nucl-ex]} \BibitemShut {NoStop}%
\bibitem [{\citenamefont {Agakishiev}\ \emph {et~al.}(2012)\citenamefont {Agakishiev} \emph {et~al.}}]{ref:starPion2012}%
  \BibitemOpen
  \bibfield  {author} {\bibinfo {author} {\bibfnamefont {G.}~\bibnamefont {Agakishiev}} \emph {et~al.} (\bibinfo {collaboration} {STAR}),\ }\href {\doibase 10.1103/PhysRevLett.108.072302} {\bibfield  {journal} {\bibinfo  {journal} {Phys. Rev. Lett.}\ }\textbf {\bibinfo {volume} {108}},\ \bibinfo {pages} {072302} (\bibinfo {year} {2012})},\ \Eprint {http://arxiv.org/abs/1110.0579} {arXiv:1110.0579 [nucl-ex]} \BibitemShut {NoStop}%
\bibitem [{\citenamefont {Abdallah}\ \emph {et~al.}(2021{\natexlab{b}})\citenamefont {Abdallah} \emph {et~al.}}]{ref:STAR_invariant_mass}%
  \BibitemOpen
  \bibfield  {author} {\bibinfo {author} {\bibfnamefont {M.}~\bibnamefont {Abdallah}} \emph {et~al.} (\bibinfo {collaboration} {STAR}),\ }\href@noop {} {\bibfield  {journal} {\bibinfo  {journal} {Phys. Rev. D}\ }\textbf {\bibinfo {volume} {104}},\ \bibinfo {pages} {052007} (\bibinfo {year} {2021}{\natexlab{b}})},\ \Eprint {http://arxiv.org/abs/2103.13286} {arXiv:2103.13286 [hep-ex]} \BibitemShut {NoStop}%
\bibitem [{\citenamefont {Adam}\ \emph {et~al.}(2020{\natexlab{a}})\citenamefont {Adam} \emph {et~al.}}]{ref:STAR_groomed_jet}%
  \BibitemOpen
  \bibfield  {author} {\bibinfo {author} {\bibfnamefont {J.}~\bibnamefont {Adam}} \emph {et~al.} (\bibinfo {collaboration} {STAR}),\ }\href@noop {} {\bibfield  {journal} {\bibinfo  {journal} {Phys. Lett. B}\ }\textbf {\bibinfo {volume} {811}},\ \bibinfo {pages} {135846} (\bibinfo {year} {2020}{\natexlab{a}})},\ \Eprint {http://arxiv.org/abs/2003.02114} {arXiv:2003.02114 [hep-ex]} \BibitemShut {NoStop}%
\bibitem [{\citenamefont {Adam}\ \emph {et~al.}(2020{\natexlab{b}})\citenamefont {Adam} \emph {et~al.}}]{ref:STAR_UE_paper}%
  \BibitemOpen
  \bibfield  {author} {\bibinfo {author} {\bibfnamefont {J.}~\bibnamefont {Adam}} \emph {et~al.} (\bibinfo {collaboration} {STAR}),\ }\href {\doibase 10.1103/PhysRevD.101.052004} {\bibfield  {journal} {\bibinfo  {journal} {Phys. Rev. D}\ }\textbf {\bibinfo {volume} {101}},\ \bibinfo {pages} {052004} (\bibinfo {year} {2020}{\natexlab{b}})},\ \Eprint {http://arxiv.org/abs/1912.08187} {arXiv:1912.08187 [nucl-ex]} \BibitemShut {NoStop}%
\bibitem [{\citenamefont {Abdallah}\ \emph {et~al.}(2022{\natexlab{b}})\citenamefont {Abdallah} \emph {et~al.}}]{ref:STAR_2012_2015_Collins}%
  \BibitemOpen
  \bibfield  {author} {\bibinfo {author} {\bibfnamefont {M.}~\bibnamefont {Abdallah}} \emph {et~al.} (\bibinfo {collaboration} {STAR}),\ }\href {\doibase 10.1103/PhysRevD.106.072010} {\bibfield  {journal} {\bibinfo  {journal} {Phys. Rev. D}\ }\textbf {\bibinfo {volume} {106}},\ \bibinfo {pages} {072010} (\bibinfo {year} {2022}{\natexlab{b}})},\ \Eprint {http://arxiv.org/abs/2205.11800} {arXiv:2205.11800 [hep-ex]} \BibitemShut {NoStop}%
\bibitem [{\citenamefont {Brun}\ \emph {et~al.}(1987)\citenamefont {Brun}, \citenamefont {Bruyant}, \citenamefont {Maire}, \citenamefont {McPherson},\ and\ \citenamefont {Zanarini}}]{ref:Geant}%
  \BibitemOpen
  \bibfield  {author} {\bibinfo {author} {\bibfnamefont {R.}~\bibnamefont {Brun}}, \bibinfo {author} {\bibfnamefont {F.}~\bibnamefont {Bruyant}}, \bibinfo {author} {\bibfnamefont {M.}~\bibnamefont {Maire}}, \bibinfo {author} {\bibfnamefont {A.~C.}\ \bibnamefont {McPherson}}, \ and\ \bibinfo {author} {\bibfnamefont {P.}~\bibnamefont {Zanarini}},\ }\href@noop {} {\bibfield  {journal} {\bibinfo  {journal} {GEANT3 Report No. CERN-DD-EE-84-1}\ } (\bibinfo {year} {1987})},\ \Eprint {http://arxiv.org/abs/http://inspirehep.net/record/252007} {http://inspirehep.net/record/252007} \BibitemShut {NoStop}%
\bibitem [{\citenamefont {Cacciari}\ \emph {et~al.}(2012)\citenamefont {Cacciari}, \citenamefont {Salam},\ and\ \citenamefont {Soyez}}]{ref:FastJet}%
  \BibitemOpen
  \bibfield  {author} {\bibinfo {author} {\bibfnamefont {M.}~\bibnamefont {Cacciari}}, \bibinfo {author} {\bibfnamefont {G.~P.}\ \bibnamefont {Salam}}, \ and\ \bibinfo {author} {\bibfnamefont {G.}~\bibnamefont {Soyez}},\ }\href {\doibase 10.1140/epjc/s10052-012-1896-2} {\bibfield  {journal} {\bibinfo  {journal} {Eur. Phys. J. C}\ }\textbf {\bibinfo {volume} {72}},\ \bibinfo {pages} {1896} (\bibinfo {year} {2012})},\ \Eprint {http://arxiv.org/abs/1111.6097} {arXiv:1111.6097 [hep-ph]} \BibitemShut {NoStop}%
\bibitem [{\citenamefont {Frixione}\ and\ \citenamefont {Ridolfi}(1997{\natexlab{a}})}]{Frixione:1997ks}%
  \BibitemOpen
  \bibfield  {author} {\bibinfo {author} {\bibfnamefont {S.}~\bibnamefont {Frixione}}\ and\ \bibinfo {author} {\bibfnamefont {G.}~\bibnamefont {Ridolfi}},\ }\href {\doibase 10.1016/S0550-3213(97)00575-0} {\bibfield  {journal} {\bibinfo  {journal} {Nucl. Phys. B}\ }\textbf {\bibinfo {volume} {507}},\ \bibinfo {pages} {315} (\bibinfo {year} {1997}{\natexlab{a}})},\ \Eprint {http://arxiv.org/abs/hep-ph/9707345} {arXiv:hep-ph/9707345} \BibitemShut {NoStop}%
\bibitem [{\citenamefont {Bouchet}(2009)}]{ref:STAR_HFT}%
  \BibitemOpen
  \bibfield  {author} {\bibinfo {author} {\bibfnamefont {J.}~\bibnamefont {Bouchet}},\ }\href {\doibase 10.1016/j.nuclphysa.2009.10.114} {\bibfield  {journal} {\bibinfo  {journal} {Nucl. Phys. A}\ }\textbf {\bibinfo {volume} {830}},\ \bibinfo {pages} {636c} (\bibinfo {year} {2009})},\ \Eprint {http://arxiv.org/abs/0907.3407} {arXiv:0907.3407 [nucl-ex]} \BibitemShut {NoStop}%
\bibitem [{\citenamefont {Cacciari}\ and\ \citenamefont {Salam}(2008)}]{ref:Ghost_Particle}%
  \BibitemOpen
  \bibfield  {author} {\bibinfo {author} {\bibfnamefont {M.}~\bibnamefont {Cacciari}}\ and\ \bibinfo {author} {\bibfnamefont {G.~P.}\ \bibnamefont {Salam}},\ }\href {\doibase 10.1016/j.physletb.2007.09.077} {\bibfield  {journal} {\bibinfo  {journal} {Phys. Lett. B}\ }\textbf {\bibinfo {volume} {659}},\ \bibinfo {pages} {119} (\bibinfo {year} {2008})},\ \Eprint {http://arxiv.org/abs/0707.1378} {arXiv:0707.1378 [hep-ph]} \BibitemShut {NoStop}%
\bibitem [{\citenamefont {Voss}\ \emph {et~al.}(2007)\citenamefont {Voss}, \citenamefont {Hocker}, \citenamefont {Stelzer},\ and\ \citenamefont {Tegenfeldt}}]{ref:tmva}%
  \BibitemOpen
  \bibfield  {author} {\bibinfo {author} {\bibfnamefont {H.}~\bibnamefont {Voss}}, \bibinfo {author} {\bibfnamefont {A.}~\bibnamefont {Hocker}}, \bibinfo {author} {\bibfnamefont {J.}~\bibnamefont {Stelzer}}, \ and\ \bibinfo {author} {\bibfnamefont {F.}~\bibnamefont {Tegenfeldt}},\ }\href {\doibase 10.22323/1.050.0040} {\bibfield  {journal} {\bibinfo  {journal} {PoS}\ }\textbf {\bibinfo {volume} {ACAT}},\ \bibinfo {pages} {040} (\bibinfo {year} {2007})}\BibitemShut {NoStop}%
\bibitem [{\citenamefont {Craigie}\ \emph {et~al.}(1983)\citenamefont {Craigie}, \citenamefont {Hidaka}, \citenamefont {Jacob},\ and\ \citenamefont {Renard}}]{Craigie:1983qjl}%
  \BibitemOpen
  \bibfield  {author} {\bibinfo {author} {\bibfnamefont {N.~S.}\ \bibnamefont {Craigie}}, \bibinfo {author} {\bibfnamefont {K.}~\bibnamefont {Hidaka}}, \bibinfo {author} {\bibfnamefont {M.}~\bibnamefont {Jacob}}, \ and\ \bibinfo {author} {\bibfnamefont {F.~M.}\ \bibnamefont {Renard}},\ }\href {\doibase 10.1016/0370-1573(93)90008-2} {\bibfield  {journal} {\bibinfo  {journal} {Phys. Rept.}\ }\textbf {\bibinfo {volume} {99}},\ \bibinfo {pages} {69} (\bibinfo {year} {1983})}\BibitemShut {NoStop}%
\bibitem [{\citenamefont {Mukherjee}\ and\ \citenamefont {Vogelsang}(2012)}]{PhysRevD.86.094009}%
  \BibitemOpen
  \bibfield  {author} {\bibinfo {author} {\bibfnamefont {A.}~\bibnamefont {Mukherjee}}\ and\ \bibinfo {author} {\bibfnamefont {W.}~\bibnamefont {Vogelsang}},\ }\href {\doibase 10.1103/PhysRevD.86.094009} {\bibfield  {journal} {\bibinfo  {journal} {Phys. Rev. D}\ }\textbf {\bibinfo {volume} {86}},\ \bibinfo {pages} {094009} (\bibinfo {year} {2012})},\ \bibinfo {note} {[Erratum: Phys.Rev.D 107, 119901 (2023)]},\ \Eprint {http://arxiv.org/abs/1209.1785} {arXiv:1209.1785 [hep-ph]} \BibitemShut {NoStop}%
\bibitem [{\citenamefont {de~Florian}\ \emph {et~al.}(1999)\citenamefont {de~Florian}, \citenamefont {Frixione}, \citenamefont {Signer},\ and\ \citenamefont {Vogelsang}}]{deFlorian:1998qp}%
  \BibitemOpen
  \bibfield  {author} {\bibinfo {author} {\bibfnamefont {D.}~\bibnamefont {de~Florian}}, \bibinfo {author} {\bibfnamefont {S.}~\bibnamefont {Frixione}}, \bibinfo {author} {\bibfnamefont {A.}~\bibnamefont {Signer}}, \ and\ \bibinfo {author} {\bibfnamefont {W.}~\bibnamefont {Vogelsang}},\ }\href {\doibase 10.1016/S0550-3213(98)00673-7} {\bibfield  {journal} {\bibinfo  {journal} {Nucl. Phys. B}\ }\textbf {\bibinfo {volume} {539}},\ \bibinfo {pages} {455} (\bibinfo {year} {1999})},\ \Eprint {http://arxiv.org/abs/hep-ph/9808262} {arXiv:hep-ph/9808262} \BibitemShut {NoStop}%
\bibitem [{\citenamefont {Martin}\ \emph {et~al.}(2009)\citenamefont {Martin}, \citenamefont {Stirling}, \citenamefont {Thorne},\ and\ \citenamefont {Watt}}]{Martin:2009iq}%
  \BibitemOpen
  \bibfield  {author} {\bibinfo {author} {\bibfnamefont {A.~D.}\ \bibnamefont {Martin}}, \bibinfo {author} {\bibfnamefont {W.~J.}\ \bibnamefont {Stirling}}, \bibinfo {author} {\bibfnamefont {R.~S.}\ \bibnamefont {Thorne}}, \ and\ \bibinfo {author} {\bibfnamefont {G.}~\bibnamefont {Watt}},\ }\href {\doibase 10.1140/epjc/s10052-009-1072-5} {\bibfield  {journal} {\bibinfo  {journal} {Eur. Phys. J. C}\ }\textbf {\bibinfo {volume} {63}},\ \bibinfo {pages} {189} (\bibinfo {year} {2009})},\ \Eprint {http://arxiv.org/abs/0901.0002} {arXiv:0901.0002 [hep-ph]} \BibitemShut {NoStop}%
\bibitem [{\citenamefont {Ball}\ \emph {et~al.}(2013)\citenamefont {Ball}, \citenamefont {Bertone}, \citenamefont {Carrazza}, \citenamefont {Del~Debbio}, \citenamefont {Forte}, \citenamefont {Guffanti}, \citenamefont {Hartland},\ and\ \citenamefont {Rojo}}]{Ball:2013hta}%
  \BibitemOpen
  \bibfield  {author} {\bibinfo {author} {\bibfnamefont {R.~D.}\ \bibnamefont {Ball}}, \bibinfo {author} {\bibfnamefont {V.}~\bibnamefont {Bertone}}, \bibinfo {author} {\bibfnamefont {S.}~\bibnamefont {Carrazza}}, \bibinfo {author} {\bibfnamefont {L.}~\bibnamefont {Del~Debbio}}, \bibinfo {author} {\bibfnamefont {S.}~\bibnamefont {Forte}}, \bibinfo {author} {\bibfnamefont {A.}~\bibnamefont {Guffanti}}, \bibinfo {author} {\bibfnamefont {N.~P.}\ \bibnamefont {Hartland}}, \ and\ \bibinfo {author} {\bibfnamefont {J.}~\bibnamefont {Rojo}} (\bibinfo {collaboration} {NNPDF}),\ }\href {\doibase 10.1016/j.nuclphysb.2013.10.010} {\bibfield  {journal} {\bibinfo  {journal} {Nucl. Phys. B}\ }\textbf {\bibinfo {volume} {877}},\ \bibinfo {pages} {290} (\bibinfo {year} {2013})},\ \Eprint {http://arxiv.org/abs/1308.0598} {arXiv:1308.0598 [hep-ph]} \BibitemShut {NoStop}%
\bibitem [{\citenamefont {Altarelli}\ \emph {et~al.}(1998)\citenamefont {Altarelli}, \citenamefont {Forte},\ and\ \citenamefont {Ridolfi}}]{ref:positivity_Altarelli:1998gn}%
  \BibitemOpen
  \bibfield  {author} {\bibinfo {author} {\bibfnamefont {G.}~\bibnamefont {Altarelli}}, \bibinfo {author} {\bibfnamefont {S.}~\bibnamefont {Forte}}, \ and\ \bibinfo {author} {\bibfnamefont {G.}~\bibnamefont {Ridolfi}},\ }\href {\doibase 10.1016/S0550-3213(98)00661-0} {\bibfield  {journal} {\bibinfo  {journal} {Nucl. Phys. B}\ }\textbf {\bibinfo {volume} {534}},\ \bibinfo {pages} {277} (\bibinfo {year} {1998})},\ \Eprint {http://arxiv.org/abs/hep-ph/9806345} {arXiv:hep-ph/9806345} \BibitemShut {NoStop}%
\bibitem [{\citenamefont {de~Florian}\ \emph {et~al.}(2024)\citenamefont {de~Florian}, \citenamefont {Forte},\ and\ \citenamefont {Vogelsang}}]{deFlorian:2024utd}%
  \BibitemOpen
  \bibfield  {author} {\bibinfo {author} {\bibfnamefont {D.}~\bibnamefont {de~Florian}}, \bibinfo {author} {\bibfnamefont {S.}~\bibnamefont {Forte}}, \ and\ \bibinfo {author} {\bibfnamefont {W.}~\bibnamefont {Vogelsang}},\ }\href {\doibase 10.1103/PhysRevD.109.074007} {\bibfield  {journal} {\bibinfo  {journal} {Phys. Rev. D}\ }\textbf {\bibinfo {volume} {109}},\ \bibinfo {pages} {074007} (\bibinfo {year} {2024})},\ \Eprint {http://arxiv.org/abs/2401.10814} {arXiv:2401.10814 [hep-ph]} \BibitemShut {NoStop}%
\bibitem [{\citenamefont {Frixione}\ \emph {et~al.}(1996)\citenamefont {Frixione}, \citenamefont {Kunszt},\ and\ \citenamefont {Signer}}]{ref:Frixione:1995ms}%
  \BibitemOpen
  \bibfield  {author} {\bibinfo {author} {\bibfnamefont {S.}~\bibnamefont {Frixione}}, \bibinfo {author} {\bibfnamefont {Z.}~\bibnamefont {Kunszt}}, \ and\ \bibinfo {author} {\bibfnamefont {A.}~\bibnamefont {Signer}},\ }\href {\doibase 10.1016/0550-3213(96)00110-1} {\bibfield  {journal} {\bibinfo  {journal} {Nucl. Phys. B}\ }\textbf {\bibinfo {volume} {467}},\ \bibinfo {pages} {399} (\bibinfo {year} {1996})},\ \Eprint {http://arxiv.org/abs/hep-ph/9512328} {arXiv:hep-ph/9512328} \BibitemShut {NoStop}%
\bibitem [{\citenamefont {Frixione}\ and\ \citenamefont {Ridolfi}(1997{\natexlab{b}})}]{ref:Frixione:1997ks}%
  \BibitemOpen
  \bibfield  {author} {\bibinfo {author} {\bibfnamefont {S.}~\bibnamefont {Frixione}}\ and\ \bibinfo {author} {\bibfnamefont {G.}~\bibnamefont {Ridolfi}},\ }\href {\doibase 10.1016/S0550-3213(97)00575-0} {\bibfield  {journal} {\bibinfo  {journal} {Nucl. Phys. B}\ }\textbf {\bibinfo {volume} {507}},\ \bibinfo {pages} {315} (\bibinfo {year} {1997}{\natexlab{b}})},\ \Eprint {http://arxiv.org/abs/hep-ph/9707345} {arXiv:hep-ph/9707345} \BibitemShut {NoStop}%
\bibitem [{\citenamefont {Frixione}(1997)}]{ref:Frixione:1997np}%
  \BibitemOpen
  \bibfield  {author} {\bibinfo {author} {\bibfnamefont {S.}~\bibnamefont {Frixione}},\ }\href {\doibase 10.1016/S0550-3213(97)00574-9} {\bibfield  {journal} {\bibinfo  {journal} {Nucl. Phys. B}\ }\textbf {\bibinfo {volume} {507}},\ \bibinfo {pages} {295} (\bibinfo {year} {1997})},\ \Eprint {http://arxiv.org/abs/hep-ph/9706545} {arXiv:hep-ph/9706545} \BibitemShut {NoStop}%
\bibitem [{\citenamefont {Stump}\ \emph {et~al.}(2001)\citenamefont {Stump}, \citenamefont {Pumplin}, \citenamefont {Brock}, \citenamefont {Casey}, \citenamefont {Huston}, \citenamefont {Kalk}, \citenamefont {Lai},\ and\ \citenamefont {Tung}}]{ref:Stump:2001gu}%
  \BibitemOpen
  \bibfield  {author} {\bibinfo {author} {\bibfnamefont {D.}~\bibnamefont {Stump}}, \bibinfo {author} {\bibfnamefont {J.}~\bibnamefont {Pumplin}}, \bibinfo {author} {\bibfnamefont {R.}~\bibnamefont {Brock}}, \bibinfo {author} {\bibfnamefont {D.}~\bibnamefont {Casey}}, \bibinfo {author} {\bibfnamefont {J.}~\bibnamefont {Huston}}, \bibinfo {author} {\bibfnamefont {J.}~\bibnamefont {Kalk}}, \bibinfo {author} {\bibfnamefont {H.~L.}\ \bibnamefont {Lai}}, \ and\ \bibinfo {author} {\bibfnamefont {W.~K.}\ \bibnamefont {Tung}},\ }\href {\doibase 10.1103/PhysRevD.65.014012} {\bibfield  {journal} {\bibinfo  {journal} {Phys. Rev. D}\ }\textbf {\bibinfo {volume} {65}},\ \bibinfo {pages} {014012} (\bibinfo {year} {2001})},\ \Eprint {http://arxiv.org/abs/hep-ph/0101051} {arXiv:hep-ph/0101051} \BibitemShut {NoStop}%
\end{thebibliography}%

\end{document}